\documentclass[12pt]{article}
\usepackage{graphicx}
\usepackage{amsmath}

\newtheorem{theorem}{Theorem}[section]
\newtheorem{definition}{Definition}[section]
\newtheorem{lemma}[theorem]{Lemma}

\newtheorem{corollary}[theorem]{Corollary}
\newtheorem{remark}[theorem]{Remark}
\newenvironment{proof}[1][Proof]{\textsc{#1.} }{\ \rule{0.5em}{0.5em}}
\numberwithin{equation}{section}
\input{tcilatex}

\begin{document}

\title{Future complete U(1) symmetric Einsteinian spacetimes, the unpolarized case.}
\author{Yvonne Choquet-Bruhat}
\maketitle

\section{Introduction.}

In this paper I generalize the non linear stability theorem obtained in
collaboration with V.\ Moncrief (CB-M1, CB-M2) for vacuum Einsteinian 4 -
manifolds $(V,^{(4)}g)$ where $V=M\times R$ with $M$ a circle bundle over a
compact, orientable surface $\Sigma$ of genus greater than 1. The Lorentzian
metric $^{(4)}g$ admits a Killing symmetry along the (spacelike) circular
fibers. I remove the so-called polarization condition, i.e. the
orthogonality of the fibers to quotient 3-manifolds. The reduced field
equations take now the form of a wave map equation, instead of a linear wave
equation in the polarized case, coupled to 2+1 gravity. I use results on
wave maps from curved manifolds obtained in CB1, CB2. Like in CB-M2 we do
not restrict the conformal geometry of $\Sigma$ to avoid those regions of
Teichm\"{u}ller space for which the lowest positive eigenvalues of the
scalar Laplacian lie, in our normalization, in the gap $(0,\frac{1}{8}]$. A
consequence is that the asymptotic behaviour of the wave map field does not
exhibit a universal rate of decay but instead develops a decay rate which
depends upon the asymptotic values of the lowest eigenvalue.

Under the Kaluza-Klein reduction which one carries out in the presence of
the assumed spacelike Killing field one is first led to field equations of
the type of an Einstein - Maxwell - Jordan system on the 3 - manifold $%
\Sigma \times R$. To transform this to a more convenient Einstein - wave map
system one needs a further topological restriction on the fields allowed.
The need for this arises from considering the constraint equation for the
effective 2+1 dimensional electric type field density $\tilde{e}=e^{a}\frac{%
\partial }{\partial x^{a}}$ \ which reads $e_{,a}^{a}=0.$ On a higher genus
surface $\Sigma $ the general solution of this equation (which results from
a Hodge decomposition of one - forms on $\Sigma $) takes the form $e^{a}=$ $%
\varepsilon ^{ab}(\omega ,_{b}+h_{b})$ where $h_{b}dx^{b}$ is a harmonic one
- form on $\Sigma .$ A consistent simplification results from setting this
harmonic contribution to zero so that $e^{a}\frac{\partial }{\partial x^{a}}$
becomes expressible purely in terms of the so - called twist potential $%
\omega .$ Taken together with the norm of the ($U(1)$ generating) Killing
field $\tilde{Y}$, conveniently expessed via $\tilde{Y}.\tilde{Y}=$ $%
e^{2\gamma },$ the twist potential $\omega $ and the function $\gamma $
provide a map from $\Sigma \times R$ to $R^{2}.$ When expressed in terms of
the pair ($\gamma ,\omega )$ Einstein's equations take the form of a wave
map from a 2+1 $\ $Lorentzian manifold $(\Sigma \times R,^{(3)}g)$ into the
Poincar\'{e} plane with its standard metric $2d\gamma ^{2}$ +$\frac{1}{2}%
e^{-4\gamma }d\omega ^{2};$ the metric $^{(3)}g$ satisfies the 2+1 Einstein
equations on $\Sigma \times R$, with source the wave map. These 2+1 Einstein
equations, supplemented by suitable coordinate conditions to fix the gauge,
reduce to an elliptic system on each slice $\Sigma _{t}$ of $\Sigma \times R$
for the lapse, the shift, and the conformal factor of a 2 dimensional
metric, together with an ordinary differential system for the
Teichm\"{u}ller parameters which determine the conformal geometry. The wave
map field and the Teichm\"{u}ller parameters represent therefore the true
propagating gravitational degrees of freedom of the original problem.

The basic methods we use to prove existence for an infinite proper time
involve the construction of higher order energies to control the Sobolev
norms of the wave map and the solution of the differential system\footnote{%
{\footnotesize We use here directly this system, instead of introducing the
Dirichlet energy as ib CB-M 1 and CB6M\ 2.}} satisfied by the
Teichm\"{u}ller parameters degrees of freedom. A subtlety is that the most
obvious definition of wave map energies does not lead to a well defined rate
of decay so that suitable corrected energies must be developed which exploit
information about the lowest eigenvalues of the spatial Laplacian which
appears in the relevant wave operators. The eigenvalues vary with position
in Teichm\"{u}ller space and thus evolve along with Teichmuller parameters.
If the lowest (non trivial) eigenvalue asymptotically avoids a well known
gap in the spectrum ( the gap (0,$\frac{1}{8}]$ in our normalization which
has the more familiar form (0,$\frac{1}{4}]$ if one instead normalizes the
Gauss curvature on the higher genus surface) then we obtain a universal rate
of decay for the energies asymptotically. If the lowest eigenvalue however
drifts into this gap and remains there asymptotically then the rate of decay
of the energies will depend upon the asymptotic value of this lowest
eigenvalue and will no longer be universal. We need slightly different forms
for the corrected energies to handle these different eventualities
(universal versus non universal rates of decay). In all cases the conformal
geometry of our circle bundles undergoes a kind of Cheeger - Gromov collapse
in which the circular fibers (after a conformal rescaling needed to take out
the overall expansion) collapse to zero length asymptotically while only the
conformal 2 - geometry remains well behaved. In our set up the Sobolev
constants depend only on the conformal 2 - geometry (i.e. upon the
Teichm\"{u}ller parameters) and, so long as the evolution remains in a
compact subspace of Teichm\"{u}ller space, these constants remain under
control.

The sense in which our solutions are global in the expanding direction is
that they exhaust the maximal range allowed for the mean curvature function
on a manifold of negative Yamabe type, for which a zero mean curvature
cannot be achieved but only asymptotically approached. In addition however
our estimates prove that the normal trajectories to our spatial slices all
have infinite future proper time length, and allow us to establish, using
[CB-C], causal geodesic completeness in the expanding direction.

If the harmonic one - form $H$\ discussed above were allowed to be non zero
it would disturb the pure wave map character of the reduced field equations.
On the other hand it seems plausible that energy arguments could still be
made to work in the presence of $H.$ Alternatively one might simply refrain
from trying to force the reduced field equations into a wave map framework
and instead develop energy arguments for the Einstein - Maxwell - Jordan
type system itself which require no splitting of $\tilde{e}$ into twist
potential and harmonic contributions. I shall not however pursue either of
these possibilities here but leave them for further study.

I need as in CB-M1 and CB-M2\ a smallness condition on suitably defined
energies which control the norms of the evolving (here wave map) field and
for this reason I continue to restrict my attention to trivial $S^{1}$
bundles over $\Sigma $ (i.e., to those for which $M=S^{1}\times \Sigma $).
The reason for this is that the curvature of the $U(1)$ connection and its
assumed ($U(1)$ - generating) Killing field has a quantized integral over $%
\Sigma $ and, in the case of a non trivial bundle when this integral is not
zero, cannot be adjusted to satisfy the smallness condition needed for the
energy argument. It seems plausible that one could probably substract off
this unavoidable topological contribution to curvature and work with
suitable energies defined for the substracted fields to handle the case of
non trivial bundles but I shall not attempt to do so here.

Another approach (suggested by V. Moncrief) to treating solutions on non
trivial $S^{1}$ bundles involves applying a well known action of $SL(2,R)$
(the isometry goup of the Poincar\'{e} plane which plays the role of target
for our wave map fields) to the fields defined on the base manifold $%
\Sigma\times R.$ In certain cases this group action can be used to transform
solutions which lift to the trivial $S^{1}$ bundle over $\Sigma\times R$ to
other solutions which lift instead to another, non trivial bundle. There is
an obstruction to obtaining such solutions in this way since a certain
Casimir invariant (which is of course preserved under the group action) is
necessarily positive for solutions which lift to the trivial bundle (it can
be negative for a subset of solutions which lift to non trivial bundles).
This formulation has so far only been developed for the case of circle
bundles over $S^{2}\times R$ but can most likely be generalized to the cases
of bundles over $\Sigma\times R$ where $\Sigma$ is either a torus or a
higher genus surface. That possibility is left for further study.

The small data future global existence theorem for solutions of Einstein's
equations of Andersson and Moncrief, this volume, makes no symmetry
assumption whatsoever, but treats a different class of spatial 3 - manifolds
which are taken to be compact hyperbolic. The results of their analysis show
that the standard hyperbolic (i.e. constant negative curvature) metric on
such a manifold serves as an attractor for the conformal geometry under the
(future) Einstein flow. In other words the evolving conformal geometry has a
well behaved limit in that problem. This fact plays a crucial role in their
analysis since various Sobolev ''constants'' (which are in fact functionals
of the geometry) which are needed in the associated energy estimates are
asymptotically under control since they are tending toward their (regular)
limiting values for the hyperbolic metric. Thus the difficulty of
degenerating Sobolev constants, avoided by the introduction of a conformal 2
metric in the case of our $U(1)$ symmetry assumption, never arises in the
Andersson - Moncrief work.

Besides the fact that \ the $U(1)$ symmetric case is not included in the no
symmetry case treated by Andersson and Moncrief, an interest of the $U(1)$
case is that in our problem the number of effective spatial dimensions is
two, and also that there is no known ''physical'' reason why large data
solutions should develop singularities in the direction of cosmological
expansion. Black hole formation seems to be suppressed by the topological
character of the assumed Killing symmetry (which is of translational rather
than rotational type and excludes the appearance of an axis of symmetry) and
the big bang singularity is avoided by considering the future evolution from
an initially expanding Cauchy hypersurface. Any possible big crunch is
excluded by our requirement that the spatial manifold $M$ is of negative
Yamabe type (which is true of all circle bundles over higher genus
manifolds). Such manifolds are incompatible (in the vacuum and electrovacuum
cases for example) with the development of a maximal hypersurface which
would be a necessary prelude to the ''recollapse'' of an expanding universe
towards a hypothetical big crunch singularity. At a maximal hypersurface the
scalar curvature of $M$ would have to be everywhere positive - an
impossibility on any manifold of negative Yamabe type. Thus it is
conceivable that for large data future global existence holds for our
problem. Up to now the only large data global results require simplifying
assumptions so stringent that they effectively reduce the number of spatial
dimensions to one (e.g., Gowdy models and their generalizations, plane
symmetric gravitational vaves, spherically symmetric matter coupled with
gravity) or zero (e.g. Bianchi models, 2+1 gravity). Unfortunately we have
at present no way of proving this global existence, even in the polarized
case for which the wave map equation reduces to a wave equation for a scalar
function, because the reduced field equations are non local in character.
The ''background'' spacetime on which the scalar field evolves is not given
a priori but is instead a certain functional (obtained by solution of
elliptic equations) of the evolving field (and the Teichm\"{u}ller
parameters) itself. In the unpolarized case, the problem of global existence
of strong solutions for wave maps on a fixed background in 2+1 dimensions is
still unsolved. However there is a proof [M-S] of global existence of a weak
solution (with no uniqueness) for wave maps from Minkowski spacetime. Any
progress on the large data global existence, even of weak solutions, for the 
$U(1)$ - symmetric problem would represent a ''quantum jump'' forward in our
understanding of long time existence problems for Einstein's equations.

It is worth mentioning here that, again with suitable topological
restrictions, using the reduction obtained by V.\ Moncrief [M2], an
analogous Einstein - wave map form of the reduced field equations can be
obtained even when one begins with the full Einstein - Maxwell system in 3+1
dimensions.

Some steps of the proof given here have been obtained independently, using
other notations, by V.\ Moncrief. I thank him for communicating his
manuscript to me, and for numerous conversations on the subject.

\section{$S^{1}$ invariant einsteinian universes.}

\subsection{Definition.}

The spacetime manifold $V$ is a principal fiber bundle with Lie group $S^{1}$
and base $\Sigma\times R$, with $\Sigma$ a smooth orientable 2 dimensional
manifold which we suppose here to be compact and of genus greater than one.

The spacetime metric $^{(4)}g$ is invariant under the action of $S^{1}$, the
orbits are the fibers of $V$ and are supposed to be space like. We write it
in the form adapted to the bundle structure\footnote{{\footnotesize See for
instance CB-DM Kaluza Klein theories p.286. The Lorentzian metric on the
base manifold \ }$\Sigma \times R${\footnotesize \ is weighted by\ e}$%
^{-3\gamma }$\ {\footnotesize \ in order to obtain equations which split in
a nice hyperbolic - elliptic coupled form (see M1, CB-M3).}}: 
\begin{equation*}
^{(4)}g=e^{-2\gamma (3)}g+e^{2\gamma }(\theta )^{2},
\end{equation*}
where $\gamma $ and $^{(3)}g$ can be identified respectively with a scalar
function and a lorentzian metric on the base manifold $\Sigma \times R.$ In
coordinates $(x^{3},x^{\alpha })$ adapted to a local trivialization of the
bundle, with $x^{3}$ a coordinate on $S^{1}$ (i.e. with $x^{3}=0$ and $%
x^{3}=2\pi $ identified) and ($x^{\alpha })\equiv (x^{a},t),$ $a=1,2$
coordinates on $\Sigma \times R,$ it holds that: 
\begin{equation*}
^{(3)}g=-N^{2}dt^{2}+g_{ab}(dx^{a}+\nu ^{a}dt)(dx^{b}+\nu ^{b}dt)
\end{equation*}
equivalently, in terms of a moving frame 
\begin{equation*}
^{(3)}g=-N^{2}(\theta ^{0})^{2}+g_{ab}\theta ^{a}\theta ^{b},\text{ \ }%
\theta ^{0}\equiv dt,\text{\ \ }\theta ^{a}\equiv dx^{a}+\nu ^{a}dt
\end{equation*}
where $N$ and $\nu $ are respectively the lapse and shift of $^{(3)}g,$ and 
\begin{equation*}
g=g_{ab}dx^{a}dx^{b}
\end{equation*}
is a riemannian metric on $\Sigma $, depending on $t$. The 1-form $\theta ,$ 
$S^{1}$ connection on the fiber bundle $V$, is represented by 
\begin{equation*}
\theta =dx^{3}+A_{\alpha }dx^{\alpha }
\end{equation*}
The 1-form $A\equiv A_{\alpha }dx^{\alpha }$ depends on the trivialization
of $V,$ it is only a locally defined 1-form on $\Sigma \times R$ if the
bundle is not trivial.

\subsection{Twist potential.}

\subsubsection{Definition.}

The curvature of the connection locally represented by $A$ is a 2-form $F$
on $\Sigma \times R$, given by, if the equations $^{(4)}R_{\alpha 3}$=$0$
are satisfied, 
\begin{equation*}
F_{\alpha \beta }=(1/2)e^{-4\gamma }\eta _{\alpha \beta \lambda }E^{\lambda }
\end{equation*}
with $\eta $ the volume form of the metric $^{(3)}g,$ and $E$ an arbitrary
closed 1-form. Hence if $\Sigma $ is compact 
\begin{equation*}
E=d\omega +H
\end{equation*}
where $\omega $ is a scalar function on $V$, called the twist potential, and 
$H$ a representative of the 1-cohomology class of $\Sigma \times R$, for
instance defined by a 1-form on $\Sigma $, harmonic for some given
riemannian metric $m$. For simplicity we take $H=0.$

\subsubsection{Construction \ of $A.$}

The connection 1- form $\theta$ can be constructed, when $F$ is known if the
following integrability condition is satisfied (see [M1], [CB-M3]) 
\begin{equation}
\int_{\Sigma_{t}}F=\frac{1}{2}\int_{\Sigma_{t}}F_{ab}dx^{a}\wedge
dx^{b}=-\int_{\Sigma_{t}}e^{-4\gamma}N^{-1}\partial_{0}\omega\mu_{g}=2\pi n.
\end{equation}
where $n$ is the Chern number of the bundle over $\Sigma\times R.$

We will suppose here that this bundle is trivial, i.e. $n=0;$ this value of $%
n$ is the only one compatible with the smallness assumptions on the energy
that we will make. The 1 - form $A\equiv A_{\alpha}dx^{\alpha}$ is then
defined globally on $\Sigma\times R$. It satisfies the equation 
\begin{equation}
dA=F
\end{equation}
We denote by $\tilde{A}$ and $\tilde{F}$ the $t$ dependent 1 - form and 2 -
form on $\Sigma$ given by 
\begin{equation}
\tilde{A}\equiv A_{a}dx^{a},\text{ \ \ \ }\tilde{F}\equiv\frac{1}{2}%
F_{ab}dx^{a}\wedge dx^{b}.
\end{equation}
The equation 2.2 splits into 
\begin{equation}
d\tilde{A}=\tilde{F},\text{ \ \ \ \ i.e. \ \ }\partial_{a}A_{b}-\partial
_{b}A_{a}=F_{ab}
\end{equation}
and, denoting by $F_{(t)}$ the 1-form $F_{ta}dx^{a}:$ 
\begin{equation}
\partial_{t}A_{a}-\partial_{a}A_{t}=F_{ta},\text{ \ \ i.e. \ }\partial _{t}%
\tilde{A}-dA_{t}=F_{(t)}.
\end{equation}
We solve 2.4\ by introducing a smooth $m$ metric on $\Sigma.$ We denote by $%
\delta_{m}$ and $\tilde{\Delta}_{m}\equiv\delta_{m}d+d\delta_{m}$ the
codifferential and the de Rham Laplace operator in this metric. If we
suppose that $\tilde{A}$ satisfies the Coulomb gauge condition: 
\begin{equation}
\delta_{m}\tilde{A}=0.
\end{equation}
The equations 2.4 and 2.6 imply that 
\begin{equation}
\tilde{\Delta}_{m}\tilde{A}=\delta_{m}\tilde{F}.
\end{equation}
The general solution of this equation is the sum of the unique solution $%
\hat{A}$ which is $L^{2}(m)-$ orthogonal to the elements $H_{(i)}$ of a
basis of harmonic 1-forms, and an arbitrary harmonic 1-form, that is: 
\begin{equation}
\tilde{A}=\hat{A}+\sum_{i}c_{i}H_{(i)},\text{ \ \ }(\hat{A}%
,H_{(i)})_{L^{2}(m)}=0
\end{equation}
where the $c_{i}$ are $t$-dependent numbers. The solution $\hat{A}$
satisfies a Sobolev inequality 
\begin{equation}
||\hat{A}||_{H_{2}(m)}\leq C_{m}||\delta_{m}\tilde{F}||_{L^{2}(m)},\text{ \
\ }||\delta_{m}\tilde{F}||_{L^{2}(m)}\leq C_{m}||\tilde{F}||_{H_{1}(m)}
\end{equation}
A solution $\tilde{A}$ of 2.7\ satisfies 2.4 and 2.6 because 2.7 implies 
\begin{equation}
d\tilde{\Delta}_{m}\tilde{A}\equiv\tilde{\Delta}_{m}d\tilde{A}=d\delta _{m}%
\tilde{F}\equiv\tilde{\Delta}_{m}\tilde{F}
\end{equation}
since $\tilde{F}$ is closed, and 
\begin{equation}
\delta_{m}\tilde{\Delta}_{m}\tilde{A}\equiv\tilde{\Delta}_{m}\delta_{m}%
\tilde{A}=0.
\end{equation}
2.10 implies that $d\tilde{A}-\tilde{F}$ is a harmonic 2-form on $\Sigma,$
it is zero because its period, i.e. its integral on the unique 2 cycle $%
\Sigma,$ is zero (equation 2.1). 2.11 implies that the scalar function $%
\delta _{m}\tilde{A}$ is harmonic, but its integral on $\Sigma$ is zero,
since it is a divergence, therefore $\delta_{m}\tilde{A}=0.$

The equation 2.5 can be satisfied by choice of $A_{t},$ a $t$ dependent
function on $\Sigma,$ if the 1-form $\partial_{t}A-F_{(t)}$ is an exact
differential. The commutation of partial derivatives and the closure of $F$
show that a solution $\tilde{A}$ of 2.4 satisfies the equation 
\begin{equation}
\partial_{a}\partial_{t}A_{b}-\partial_{b}\partial_{t}A_{a}=%
\partial_{t}F_{ab}=\partial_{a}F_{tb}-\partial_{b}F_{ta},\text{ \ i.e. \ \ }%
d(\partial _{t}\tilde{A}-F_{(t)})=0.
\end{equation}
Since the form $\partial_{t}\tilde{A}-F_{(t)}$ is closed it will be the
differential of a function $A_{t}$ if and only if it is $L^{2}(m)$
orthogonal to the harmonic 1-forms, that is, because $\partial_{t}\hat{A}$
is like $\hat{A}$ $L^{2}(m)$ orthogonal to the $H_{(i)}$ which do not depend
on $t:$ 
\begin{equation}
(\sum_{j}\frac{dc_{j}}{dt}H_{(j)}-F_{(t)},H_{(i)})_{L^{2}(m)}=0.
\end{equation}
Choosing the $H_{(i)}^{\prime}s$ to be $L^{2}(m)$ orthonormal, this equation
reduces to: 
\begin{equation}
\frac{dc_{i}}{dt}=(F_{(t)},H_{(i)})_{L^{2}(m)}.
\end{equation}
These equations determine $c_{i}$ by integration on $t$ through its its
initial value $c_{i}(t_{0})$

We complete the determination of the scalar function $A_{t}$ by remarking
that for such a function the equation 2.5 implies\footnote{{\footnotesize %
For a scalar function it holds that }$\delta _{m}f\equiv ${\footnotesize 0
and }$\tilde{\Delta}_{m}f\equiv \Delta _{m}f.$}, using 2.6, 
\begin{equation}
\Delta _{m}A_{t}=-\delta _{m}F_{(t)},
\end{equation}
an equation which determines uniquely $A_{t}$ if we impose that its integral
on $\Sigma _{t}$ is zero. It satisfies then the inequality 
\begin{equation}
||A_{t}||_{H_{2}(m)}\leq C_{m}||\delta _{m}F_{(t)}||_{L^{2}(m)},\text{ \ \ }%
||\delta _{m}F_{(t)}||_{L^{2}(m)}\leq C_{m}||F_{t}||_{H_{1}(m)}.
\end{equation}

\begin{remark}
We can, instead of the Coulomb gauge, determine $\tilde{A}$ in temporal
gauge, i.e. impose $A_{t}=0.$ We determine the 1 - form $\tilde{A}_{0}$, the
value of $\tilde{A}$ for $t=t_{0}$ by the relation 2.4 through the value of $%
\tilde{F}_{0}$ as above. The equation 2.5 is, when $A_{t}$ and $F_{(t)}$ are
known, an ordinary differential equation for $\tilde{A}$ which can be solved
by integration on $t:$%
\begin{equation}
\tilde{A}=\tilde{A}_{0}+\int_{t_{0}}^{t}F_{(t)}.
\end{equation}
When 2.5 is satisfied it implies, whatever $A_{t}$ may be using the
commutation of partial derivatives and the closure of $F$ the equation 
\begin{equation}
\partial _{t}(\partial _{a}A_{b}-\partial _{b}A_{a})=\partial
_{a}F_{tb}-\partial _{b}F_{ta}=\partial _{t}F_{ab}
\end{equation}
i.e. 
\begin{equation}
\partial _{t}(d\tilde{A}-\tilde{F})=0
\end{equation}
hence $d\tilde{A}-\tilde{F}=0$ for all $t$ if it is so for $t=t_{0}.$

The disavantage of the temporal gauge is that it gives only $H_{1}$
estimates for $\tilde{A}.$
\end{remark}

\subsection{Wave map equation.}

The fact that $F$ is a closed form together with the equation $%
^{(4)}R_{33}=0 $ imply (with the choice $H=0$ in the definition of $\omega)$
that the pair $u\equiv(\gamma,\omega)$ satisfies a wave map equation from ($%
\Sigma\times R,^{(3)}g)$ into the Poincar\'{e} plane $(R^{2},G),$

\begin{equation*}
G=2(d\gamma)^{2}+(1/2)e^{-4\gamma}(d\omega)^{2},
\end{equation*}
It is a system of hyperbolic type when $^{(3)}g$ is a known lorentzian
metric which reads, denoting by $^{(3)}\nabla_{\alpha}$ the components of
covariant derivatives of tensors on $\Sigma\times R$ in the metric $^{(3)}g$
in the moving frame ($\theta^{a},dt)$: 
\begin{equation*}
^{(3)}\nabla^{\alpha}\partial_{\alpha}\gamma+\frac{1}{2}e^{-4\gamma}g^{%
\alpha\beta}\partial_{\alpha}\omega\partial_{\beta}\omega=0
\end{equation*}
\begin{equation*}
^{(3)}\nabla^{\alpha}\partial_{\alpha}\omega-4g^{\alpha\beta}\partial_{%
\alpha }\omega\partial_{\beta}\gamma=0
\end{equation*}

The integral 2.1\ is independent of $t$ if $F$ is closed, hence if the wave
map equation is satisfied.

\begin{remark}
The non zero Christoffel symbols of the metric $G$ are 
\begin{equation*}
G_{22}^{1}\equiv G_{\omega\omega}^{\gamma}=\frac{1}{2}e^{-4\gamma},\text{ \
\ \ }G_{12}^{2}=G_{21}^{2}=G_{\gamma\omega}^{\omega}=-2
\end{equation*}
The scalar and Riemann curvature are: 
\begin{equation*}
R_{12,12}=-2e^{-4\gamma},\text{ \ \ }R=-4
\end{equation*}
\end{remark}

\subsection{3-dimensional Einstein equations}

When $^{(4)}R_{3\alpha}=0$ and $^{(4)}R_{33}=0$ the Einstein equations $%
^{(4)}R_{\alpha\beta}=0$ are equivalent to Einstein equations on the
3-manifold $\Sigma\times R$ for the metric $^{(3)}g$ with source the stress
energy tensor of the wave map: 
\begin{equation*}
^{(3)}R_{\alpha\beta}=\rho_{\alpha\beta}\equiv\partial_{\alpha}u.\partial
_{\beta}u
\end{equation*}
where a dot denotes a scalar product in the metric of the Poincar\'{e}
plane: 
\begin{equation*}
\partial_{\alpha}u.\partial_{\beta}u\equiv2\partial_{\alpha}\gamma
\partial_{\beta}\gamma+\frac{1}{2}e^{-4\gamma}\partial_{\alpha}\omega
\partial_{\beta}\omega
\end{equation*}

In dimension 3 the Einstein equations are non dynamical, except for the
conformal class of $g$ determined by Teichm\"{u}ller parameters. They
decompose into:

a. Constraints.

b. Equations for lapse and shift to be satisfied on each $\Sigma_{t}$. These
equations, as well as the constraints, are of elliptic type.

c. Evolution equations for the Teichm\"{u}ller parameters, ordinary
differential equations.

\subsubsection{ Constraints on $\Sigma_{t}$.}

One denotes by $k$ the extrinsic curvature of $\Sigma _{t}$ as submanifold
of ($\Sigma \times R,^{(3)}g).$ Then, with $\nabla $ the covariant
derivative in the metric $g,$%
\begin{equation*}
k_{ab}\equiv (2N)^{-1}(-\partial _{t}g_{ab}+\nabla _{a}\nu _{b}+\nabla
_{b}\nu _{a}),
\end{equation*}
the equations (momentum constraint) 
\begin{equation}
^{(3)}R_{0a}\equiv N(-\nabla _{b}k_{a}^{b}+\partial _{a}\tau )=\partial
_{0}u.\partial _{a}u
\end{equation}
and (hamiltonian constraint, $^{(3)}S_{00}\equiv ^{(3)}R_{00}+\frac{1}{2}%
N^{2}$ $^{(3)}R$) 
\begin{equation}
2N^{-2(3)}S_{00}\equiv R(g)-k_{b}^{a}k_{a}^{b}+\tau ^{2}=N^{-2}\partial
_{0}u.\partial _{0}u+g^{ab}\partial _{a}u.\partial _{b}u
\end{equation}
do not contain second derivatives transversal to $\Sigma _{t}$ of $g$ or $u.$
They are the constraints. To transform the constraints into an elliptic
system one uses the conformal method. We set 
\begin{equation*}
g_{ab}=e^{2\lambda }\sigma _{ab},
\end{equation*}
where $\sigma $ is a riemannian metric on $\Sigma $, depending on t, on
which we will comment later, and 
\begin{equation*}
k_{ab}=h_{ab}+\frac{1}{2}g_{ab}\tau
\end{equation*}
where $\tau $ is the $g$-trace of $k$, hence $h$ is traceless.

We denote by $D$ a covariant derivation in the metric $\sigma$. We set 
\begin{equation*}
u^{\prime}=N^{-1}\partial_{0}u
\end{equation*}
with $\partial_{0}$ the Pfaff derivative of $u$, namely 
\begin{equation*}
\partial_{0}=\frac{\partial}{\partial t}-\nu^{a}\partial_{a}\text{ with }%
\partial_{a}=\frac{\partial}{\partial x^{a}}
\end{equation*}
and 
\begin{equation*}
\overset{.}{u}=e^{2\lambda}u^{\prime}
\end{equation*}

The momentum constraint on $\Sigma_{t}$ reads if $\tau$ is constant in
space, a choice which we will make 
\begin{equation}
D_{b}h_{a}^{b}=L_{a},L_{a}\equiv-D_{a}u.\overset{.}{u}
\end{equation}
This is a linear equation for $h,$ with left hand side independent of $%
\lambda.$ The general solution is the sum of a transverse traceless tensor $%
h_{TT}\equiv q$ (see 2.28 below) and a conformal Lie derivative $r.$ Such
tensors are $L^{2}-$orthogonal on $(\Sigma,\sigma).$

The hamiltonian constraint reads as the semilinear elliptic equation in $%
\lambda:$ 
\begin{equation}
\Delta\lambda=f(x,\lambda)\equiv p_{1}e^{2\lambda}-p_{2}e^{-2\lambda}+p_{3},
\end{equation}
with $\Delta\equiv\Delta_{\sigma}$ the Laplacian in the metric $\sigma$ and: 
\begin{equation*}
p_{1}\equiv\frac{1}{4}\tau^{2},\text{ \ \ }p_{2}\equiv\frac{1}{2}(\mid 
\overset{.}{u}\mid^{2}+\mid h\mid^{2}),\text{ \ \ }p_{3}\equiv\frac{1}{2}%
(R(\sigma)-|Du|^{2})
\end{equation*}

\subsubsection{ Equations for lapse and shift.}

Lapse and shift are gauge parameters for which we obtain elliptic equations
on each $\Sigma_{t}$ as follows.

We impose that the $\Sigma_{t}^{\prime}s$ have constant (in space) mean
curvature, namely that $\tau$ is a given increasing function of $t$. The
lapse $N$ satisfies then the linear elliptic equation 
\begin{equation}
\Delta N-\alpha N=-e^{2\lambda}\frac{\partial\tau}{\partial t}
\end{equation}
with (\TEXTsymbol{\vert}.\TEXTsymbol{\vert}pointwise norm in the metric $%
\sigma)$ 
\begin{equation}
\alpha\equiv e^{-2\lambda}(\mid h\mid^{2}+\mid\overset{.}{u}\mid^{2})+\frac
{1}{2}e^{2\lambda}\tau^{2}
\end{equation}
The equation to be satisfied by the shift $\nu$ results from the the
expression for $h$ deduced from the definition of $k$%
\begin{equation*}
h_{ab}\equiv(2N)^{-1}[-(\partial_{t}g_{ab}-\frac{1}{2}g_{ab}g^{cd}\partial
_{t}g_{cd})+\nabla_{a}\nu_{b}+\nabla_{b}\nu_{a}-g_{ab}\nabla_{c}\nu^{c}]
\end{equation*}
which implies, if $g_{ab}\equiv e^{2\lambda}\sigma_{ab}$ and if $n_{a}\equiv
e^{-2\lambda}\nu_{a}$ denotes the covariant components of the shift vector $%
\nu$ in the metric $\sigma$ (thus $n^{a}=\nu^{a})$%
\begin{equation*}
h_{ab}\equiv(2N)^{-1}e^{2\lambda}[-(\partial_{t}\sigma_{ab}-\frac{1}{2}%
\sigma_{ab}\sigma^{cd}\partial_{t}\sigma_{cd})+D_{a}n_{b}+D_{b}n_{a}-%
\sigma_{ab}D_{c}n^{c}]
\end{equation*}
The equation is therefore:

\begin{equation}
(L_{\sigma}n)_{ab}\equiv D_{a}n_{b}+D_{b}n_{a}-\sigma_{ab}D_{c}n^{c}=f_{ab}%
\text{ \ \ }
\end{equation}
\begin{equation}
f_{ab}\equiv2Ne^{-2\lambda}h_{ab}+\partial_{t}\sigma_{ab}-\frac{1}{2}%
\sigma_{ab}\sigma^{cd}\partial_{t}\sigma_{cd}
\end{equation}
The homogeneous associated operator, the conformal Killing operator $%
L_{\sigma},$ has injective symbol, and it has a kernel zero, since manifolds
of genus greater than 1 admit no conformal Killing fields.

The kernel of the dual of $L_{\sigma}$ is the space of transverse traceless
symmetric 2-tensors, i.e. symmetric 2-tensors $T$ such that

\begin{equation}
\sigma^{ab}T_{ab}=0,\text{ \ \ }D^{a}T_{ab}=0.
\end{equation}
These tensors are usually called TT tensors. The spaces of TT tensors are
the same for two conformal metrics.

\subsection{ Teichm\"{u}ller parameters.}

On a compact 2-dimensional manifold of genus G $\geq 2$ the space $T_{eich}$
of conformally inequivalent riemannian metrics, called Teichm\"{u}ller
space, can be identified ( Fisher and Tromba, see [F-T] or [CB-DM]) with $%
M_{-1}/D_{0}$, the quotient of the space of metrics with scalar curvature $%
-1 $ by the group of diffeomorphisms homotopic to the identity. $M_{-1}%
\mathcal{\rightarrow }T_{eich}$ is a trivial fiber bundle whose base can be
endowed with the structure of the manifold $R^{n}$, with $n=6G-6$.

We require the metric $\sigma_{t}$ to be in some chosen cross section $%
Q\rightarrow\psi(Q)$ of the above fiber bundle. Let $Q^{I},I=1,...,n$ be
coordinates in $T_{eich}$, then $\partial\psi/\partial Q^{I}$ is a known
tangent vector to $M_{-1}$ at $\psi(Q)$, that is a symmetric 2-tensor field
on $\Sigma,$ the sum of a transverse traceless tensor field $X_{I}(Q)$ and
of the Lie derivative of a vector field on the manifold $(\Sigma,\psi(Q))$.
The tensor fields $X_{I}(Q),I=1,...n$ span the space of transverse traceless
tensor fields on $(\Sigma,\psi(Q)).$ The matrix with elements 
\begin{equation*}
\int_{\Sigma}X_{I}^{ab}X_{Jab}\mu_{\psi(Q)}
\end{equation*}
is invertible.

We have found in [CB-M1] an ordinary differential system satisfied by $%
t\mapsto Q(t)$ by using on the one hand the solvability condition for the
shift equation which determines $dQ^{I}/dt$ in terms of $h_{t}$ which reads 
\begin{equation}
\int_{\Sigma_{t}}f_{ab}X_{J}^{ab}\mu_{\sigma_{t}}=0,\text{ \ }J=1,...6G-6,
\end{equation}
and on the other hand the necessary and sufficient conditions for the
previous equations to imply also the remaining equations $%
^{(3)}R_{ab}-\rho_{ab}=0,$ that is: 
\begin{equation}
\int_{\Sigma_{t}}N(^{(3)}R_{ab}-\rho_{ab})X_{J}^{ab}\mu_{\sigma_{t}}=0,\text{
for }J=1,2,...6G-6.
\end{equation}

We have used the expression 
\begin{equation*}
\partial_{t}\sigma_{ab}=\frac{dQ^{I}}{dt}X_{I,ab}+C_{ab}
\end{equation*}
where $C_{ab}$ is a Lie derivative, $L^{2}$ orthogonal to TT tensors,
together with the decomposition $h=q+r,$ with $r$ a tensor in the range of
the conformal Killing operator and $q$ a TT tensor. This last tensor can be
written with the use of the basis $X_{I}$ of such tensors, the coefficients $%
P^{I}$ depending only on $t$: 
\begin{equation*}
q_{ab}=P^{I}(t)X_{I,ab}
\end{equation*}
The orthogonality condition 2.29 reads, using the expression 2.27\ of $%
f_{ab} $ and the fact that the transverse tensors $X_{I}$ are orthogonal to
Lie derivatives and are traceless: 
\begin{equation*}
\int_{\Sigma_{t}}[2Ne^{-2%
\lambda}(r_{ab}+P^{I}X_{I,ab})+(dQ^{I}/dt)X_{I,ab}]X_{J}^{ab}\mu_{\sigma}=0
\end{equation*}
The tangent vector $dQ^{I}/dt$ to the curve $t\rightarrow Q(t)$ and the
tangent vector $P^{I}(t)$ to $T_{eich}$ are therefore linked by the linear
system 
\begin{equation*}
X_{IJ}\frac{dQ^{I}}{dt}+Y_{IJ}P^{I}+Z_{J}=0
\end{equation*}
with 
\begin{equation}
X_{IJ}\equiv\int_{\Sigma_{t}}X_{I}^{ab}X_{J,ab}\mu_{\sigma},
\end{equation}

\begin{equation}
Y_{IJ}\equiv\int_{\Sigma_{t}}2Ne^{-2\lambda}X_{I}^{ab}X_{J,ab}\mu_{\sigma },%
\text{ \ \ }Z_{J}\equiv\int_{\Sigma_{t}}2Ne^{-2\lambda}r_{ab}X_{J}^{ab}\mu_{%
\sigma}
\end{equation}
While, using 
\begin{equation}
^{(3)}R_{ab}\equiv R_{ab}-N^{-1}\overset{\_}{\partial}%
_{0}k_{ab}-2k_{ac}k_{b}^{c}+\tau k_{ab}-N^{-1}\nabla_{a}\partial_{b}N
\end{equation}
where 
\begin{equation*}
R_{ab}\equiv\frac{1}{2}Rg_{ab},\text{ \ \ }\rho_{ab}\equiv\partial
_{a}u.\partial_{b}u
\end{equation*}
\begin{equation*}
k_{ab}\equiv P^{I}X_{I,ab}+r_{ab}+\frac{1}{2}g_{ab}\tau
\end{equation*}
and $\overset{\_}{\partial}_{0}$ is an operator on time dependent space
tensors defined by, with $\mathcal{L}_{\nu}$ the Lie derivative in the
direction of $\nu,$%
\begin{equation*}
\overset{\_}{\partial}_{0}\equiv\partial_{t}-\mathcal{L}_{\nu}
\end{equation*}
gives\footnote{{\footnotesize In the formula 2.33\ indices are raised with
g, in 2.34 they are raised with }$\sigma.$} for 2.30 the expression: 
\begin{equation}
\int_{\Sigma_{t}}(-\overset{\_}{\partial}_{0}k_{ab}-2Ne^{2%
\lambda}h_{ac}h_{b}^{c}+\tau
Nh_{ab}-\nabla_{a}\partial_{b}N-\partial_{a}u.\partial
_{b}u)X_{J}^{ab}\mu_{\sigma_{t}}=0
\end{equation}
We have thus obtained an ordinary differential system of the form 
\begin{equation*}
X_{IJ}\frac{dP^{I}}{dt}+\Phi_{J}(P,\frac{dQ}{dt})=0
\end{equation*}
where $\Phi$ is a polynomial of degree 2 in $P$ and $dQ/dt$ with
coefficients depending smoothly on $Q$ and directly but continuously on $t$
through the other unknowns, namely: 
\begin{equation*}
\Phi_{J}\equiv A_{JIK}P^{I}P^{K}+B_{JIK}P^{I}\frac{dQ^{K}}{dt}%
+C_{JI}P^{I}+D_{J}
\end{equation*}
with 
\begin{equation*}
A_{JIK}\equiv\int_{\Sigma_{t}}2Ne^{2\lambda}X_{I,a}^{c}X_{K,bc}X_{J}^{ab}%
\mu_{\sigma_{t}}
\end{equation*}
\begin{equation*}
B_{JIK}\equiv\int_{\Sigma_{t}}\frac{\partial X_{I,ab}}{\partial Q^{K}}%
X_{J}^{ab}\mu_{\sigma_{t}}
\end{equation*}
\begin{equation*}
C_{JI}\equiv\int_{\Sigma_{t}}[(-\mathcal{L}_{\nu}X_{I})_{ab}+4Ne^{-2\lambda
}r_{b}^{c}X_{I,ac}-\tau NX_{I,ab}]X_{J}^{ab}\mu_{\sigma_{t}}
\end{equation*}
and, using integration by parts and the transverse property of the $X_{I%
\text{ }}$to eliminate second derivatives of $N$ (recall that $%
\nabla_{a}\partial _{b}N\equiv
D_{a}\partial_{b}N-2\partial_{a}\lambda\partial_{b}N)$ 
\begin{equation*}
D_{J}\equiv\int_{\Sigma_{t}}(-\overset{\_}{\partial}_{0}r_{ab}-2Ne^{-2%
\lambda }r_{ac}r_{b}^{c}+\tau
Nr_{ab}+2\partial_{a}\lambda\partial_{b}N-\partial
_{a}u.\partial_{b}u)X_{J}^{ab}\mu_{\sigma_{t}}.
\end{equation*}

\section{Cauchy problem.}

\subsection{Cauchy data.}

The Cauchy data on $\Sigma_{t_{0}}$ are:

1. A $C^{\infty}$ riemannian metric $\sigma_{0}$ which projects onto a point 
$Q(t_{0})$ of $T_{eich}$ and a $C^{\infty}$ tensor $q_{0}$ which is TT in
the metric $\sigma_{0}.$

2. Cauchy data for $u$ and $\dot{u}$ on $\Sigma_{t_{0}}$, i.e. 
\begin{equation*}
u(t_{0},.)=u_{0},\text{ \ \ \ }\dot{u}(t_{0},.)=\dot{u}_{0}.
\end{equation*}

We say that a pair of scalar functions, $u\equiv(\gamma,\omega)$ or $\dot
{u}\equiv(\dot{\gamma},\dot{\omega})$ belongs to $W_{s}^{p}$ if it is so of
each of the scalars; $W_{s}^{p}$ and $H_{s}\equiv W_{s}^{2}$ are the usual
Sobolev spaces of scalar functions on the riemannian manifold $(\Sigma
,\sigma_{0}).$ We suppose that 
\begin{equation*}
u_{0}\in H_{2},\text{ \ \ \ }\dot{u}_{0}\in H_{1}.
\end{equation*}
From these data one determines the values on $\Sigma_{0}$ of the auxiliary
unkown, $h_{0}\in W_{2}^{p},$ $1<p<2,$ the conformal factor lapse and shift $%
\lambda_{0},$ $N_{0},$ $\nu_{0}\in W_{3}^{p}.$

One deduces then the usual Cauchy data for the wave map by 
\begin{equation}
(\partial_{t}u)_{0}=e^{-2\lambda_{0}}N_{0}\dot{u}_{0}+\nu_{0}^{a}\partial
_{a}u_{0}
\end{equation}
It holds that 
\begin{equation}
(\partial_{t}u)_{0}\in H_{1}.
\end{equation}
We suppose that the initial data satisfy the integrability condition 2.1 and
we deduce from them an admissible $\tilde{A}_{0}.$

\subsection{Local in time existence theorem.}

The following theorem is a consequence of previous results (see CB-M2, CB1,
CB-M1).

\begin{theorem}
The Cauchy problem with the above data for the Einstein equations with $%
S^{1} $ isometry group has, if $T-t_{0\text{ }}$is small enough, a solution
with $u\in$ $C^{0}([t_{0},T),H_{2})$ $\dot{u}\in C^{1}([t_{0},T),H_{1});$ $%
\lambda,N,\nu\in C^{0}([t_{0},T),$ $W_{3}^{p})$ $\cap
C^{1}([t_{0},T),W_{2}^{p}),$ $1<p<2$ and $N>0$ while $\ \sigma\in
C^{1}([t_{0},T),C^{\infty})$ with $\sigma_{t}$ uniformly equivalent to $%
\sigma_{0}$. This solution is unique up to $t$ parametrization of $\tau,$
choice of $A_{t}$, and choice of a cross section of $M_{-1}$ over $T_{eich}.$
\end{theorem}

\subsection{Scheme for global existence.}

If the universe is expanding the mean curvature $\tau$ starts negative and
increases, the universe attains a moment of maximum expansion if it exists
up to $\tau=0.$ We choose the time parameter $t$ by requiring that 
\begin{equation}
t=-\frac{1}{\tau}.
\end{equation}
Then $t$ increases from $t_{0}>0$ to infinity when $\tau$ increases from $%
\tau_{0}<0$ to zero.

It results from the local existence theorem and a standard argument that the
solution of the Einstein equations exists on $[t_{0},\infty)$ if the curve $%
t\mapsto Q(t)$ remains in a compact subset of $T_{eich}$ and the norms $%
||\gamma(t,.),\omega(t,.)||_{H_{2}},$\ $||\partial_{t}\gamma(t,.),\partial
_{t}\omega(t,.)||_{H_{1}}$do not blow up for any finite $t.$

It will result from the following sections that these norms do not blow up
if it is so of the energies that we will now define. However this non blow
up will be proved only for small initial data and the proof of the
boundedness of $Q$ will require the consideration of corrected energies,
analogous to the corrected energies introduced in [CB-M1], but linked with
the wave map structure and more complicated to estimate.

In section 4 the first and second energies are defined. A proof is given
that the first energy is non increasing. Some preliminary properties of
gauge covariant derivatives are given, but the estimate of the second energy
is posponed after the estimates of the coefficients of the wave map equation
through the elliptic equations they satisfy.

In sections 5 and 6 we obtain these elliptic estimates for the difference of
various quantities with what will be their asymptotic value, in terms of the
energies previously defined. We choose $\sigma _{t}$ such that $R(\sigma
_{t})=-1,$ we suppose that the energies are bounded by some number H$_{E}$
and that the projection of $\sigma _{t}$ on the Teichm\"{u}ller space
remains in a compact subset. We first obtain bounds in $H_{1}$ for $h,$ and $%
\lambda $ in $H_{2},$ then bounds for $h$ in $W_{2}^{p},$ $2-N$ and $\nu $
in $W_{3}^{p},1<p<2.$ We also bound\footnote{{\footnotesize We did not need
this bound in CB-M1 due to the consideration of a special class of initial
data, whose property (equation 47) was conserved in time: this conservation
does not hold in the unpolarized case.}} $\partial _{t}\sigma _{t}.$

In section 8 we use the estimates found on the coefficients of the wave map
equation to obtain a non linear differential inequality for the second
energy. We could deduce from it, by a continuity argument, an a priori bound
for this energy also (the first energy has been shown to be non decreasing)
if we knew that the metrics $\sigma _{t}$ are all uniformly equivalent. To
obtain such a result we must prove the decay of the energies. This decay is
proved in section 9 and 10 through the introduction of modified energies.
The proof is more involved than in the unpolarized case, but follows
essentially the same lines. All the obtained estimates lead to a global
existence theorem by a continuity argument.

\section{Energies.}

\subsection{First energy.}

\subsubsection{Definition.}

We denote below by $|.|$ a norm in the metric $G$ and \TEXTsymbol{\vert}.%
\TEXTsymbol{\vert}$_{g}$ a norm in the metrics $g$ and $G,$ in particular: 
\begin{equation}
|u^{\prime }|^{2}\equiv 2(\gamma ^{\prime })^{2}+\frac{1}{2}e^{-4\gamma
}(\omega ^{\prime })^{2}\text{ , \ \ }|Du|_{g}^{2}\equiv g^{ab}(2D_{a}\gamma
D_{b}\gamma +\frac{1}{2}e^{-4\gamma }D_{a}\omega D_{b}\omega )
\end{equation}
The 2+1 dimensional Einstein equations with source the stress energy tensor
of the wave map $u$ contain the following equation (hamiltonian constraint) 
\begin{equation}
2N^{-2}(T_{00}-^{(3)}S_{00})=|u^{\prime
}|^{2}+|Du|_{g}^{2}+|k|_{g}^{2}-R(g)-\tau ^{2}=0
\end{equation}

The splitting of the covariant 2-tensor $k$ into a trace and a traceless
part: 
\begin{equation}
k_{ab}=h_{ab}+\frac{1}{2}g_{ab}\tau
\end{equation}
gives that:

\begin{equation}
|k|_{g}^{2}=g^{ac}g^{bd}k_{ab}k_{cd}=|h|_{g}^{2}+\frac{1}{2}\tau^{2}
\end{equation}
and the hamiltonian constraint equation reads 
\begin{equation}
|u^{\prime}|^{2}+|Du|_{g}^{2}+|h|_{g}^{2}=R(g)+\frac{1}{2}\tau^{2}
\end{equation}

Inspired by this equation, we define the first energy by the following
formula 
\begin{equation*}
E(t)\equiv\int_{\Sigma_{t}}(I_{0}+I_{1}+\frac{1}{2}|h|_{g}^{2})\mu_{g}
\end{equation*}
with 
\begin{align*}
I_{0} & \equiv{\frac{1}{2}}\mid u^{\prime}\mid^{2}\equiv(\gamma^{\prime
})^{2}+{\frac{1}{4}}e^{-4\gamma}(\omega^{\prime})^{2}, \\
I_{1} & \equiv{\frac{1}{2}}\mid Du\mid_{g}^{2}\equiv\mid D\gamma\mid_{g}^{2}+%
{\frac{1}{4}}e^{-4\gamma}\mid D\omega\mid_{g}^{2}
\end{align*}
that is:

\begin{equation}
E(t)\equiv\frac{1}{2}\{\parallel u^{\prime}\parallel_{g}^{2}+\parallel
Du\parallel_{g}^{2}+\parallel h\parallel_{g}^{2}\}
\end{equation}
with $\parallel.\parallel_{g}^{2}$the square of the integral in the metric $%
g $ of $|.|_{g}^{2}$. This energy is the first energy of the wave map $u$
completed by the square of the $L^{2}(g)$ norm of $h.$

\subsection{Bound of the first energy.}

The integration of the hamiltonian constraint on ($\Sigma_{t,}g)$ using the
constancy of $\tau$ and the Gauss Bonnet theorem which reads, with $\chi$
the Euler characteristic of $\Sigma$

\begin{equation}
\int_{\Sigma_{t}}R(g)\mu_{g}=4\pi\chi
\end{equation}
shows that 
\begin{equation*}
E(t)=\frac{\tau^{2}}{4}Vol_{g}(\Sigma_{t})+2\pi\chi,\text{ \ \ }%
Vol_{g}(\Sigma_{t})=\int_{\Sigma_{t}}\mu_{g}.
\end{equation*}
Recall that on a compact manifold 
\begin{equation*}
\frac{dVol_{g}\Sigma_{t}}{dt}=\frac{1}{2}\int_{\Sigma_{t}}g^{ab}\frac{%
\partial g_{ab}}{\partial t}\mu_{g}=-\tau\int_{\Sigma_{t}}N\mu_{g}.
\end{equation*}
We use the equation 
\begin{equation*}
N^{-1(3)}R_{00}\equiv\Delta_{g}N-N|k|_{g}^{2}+\partial_{t}\tau=N|u^{\prime
}|^{2}
\end{equation*}
together with the splitting of $k$ to write after integration, since $\tau$
is constant in space,

\begin{equation*}
\frac{1}{2}\tau ^{2}\int_{\Sigma _{t}}N\mu _{g}=\frac{d\tau }{dt}%
Vol_{g}(\Sigma _{t})-\int_{\Sigma _{t}}N(|h|_{g}^{2}+|u^{\prime }|^{2})\mu
_{g}
\end{equation*}
We then find as in [CB-M1] that it simplifies to: 
\begin{equation}
\frac{dE(t)}{dt}=\frac{1}{2}\tau \int_{\Sigma _{t}}(|h|_{g}^{2}+|u^{\prime
}|^{2})N\mu _{g}.
\end{equation}

We see that $E(t)$ is a non increasing\footnote{{\footnotesize The absence
of the term \TEXTsymbol{\vert}Du\TEXTsymbol{\vert}}$^{2}${\footnotesize \ \
prevents the use of this equality to obtain a decay estimate.}} function of $%
t$ if $\tau$ is negative. We remark that, due to the use of the constraints $%
DN$ does not appear in 4.8, as it would have if we had used only the wave
map energy.

We set 
\begin{equation}
\varepsilon\equiv\{E(t)\}^{\frac{1}{2}},\text{ \ \ }\varepsilon_{0}\equiv%
\{E(t_{0})\}^{\frac{1}{2}},
\end{equation}
we have proved that if $\tau\leq0$ then 
\begin{equation}
\varepsilon\leq\varepsilon_{0}.
\end{equation}

\subsection{Second energy.}

\subsubsection{notations.}

We denote by $\hat{\nabla}$ a covariant derivative in the metrics $g$ and $G$%
, for $t$ dependent sections of the fiber bundle $E^{p,q}$ with base $\Sigma 
$ and fiber $\otimes ^{p}T_{x}^{\ast }\Sigma \otimes ^{q}T_{u(x)}P,$ with $P$
the Poincar\'{e} plane. That is we set, for $\partial _{c}u^{A},$ a section
of $E^{1,1},$ 
\begin{equation}
\hat{\nabla}_{b}\partial _{c}u^{A}\equiv \partial _{b}\partial
_{c}u^{A}-\Gamma _{bc}^{a}\partial _{a}^{A}+G_{BC}^{A}\partial
_{b}u^{B}\partial _{c}u^{C}
\end{equation}
where $\Gamma _{bc}^{a}$ and $G_{BC}^{A}$ denote respectively the connection
coefficients of the metric $g,$ and of $G$ given in the remark 2.2. For $%
u^{\prime A},$ section of $E^{0,1},$ we have: 
\begin{equation}
\hat{\nabla}_{a}u^{\prime A}=\partial _{a}u^{\prime A}+G_{BC}^{A}\partial
_{a}u^{B}u^{\prime C}.
\end{equation}
while for $G_{AB},$ section of $E^{0,2}$ it holds that 
\begin{equation}
\hat{\nabla}_{a}G_{AB}=0.
\end{equation}
On the other hand we define by $\hat{\partial}_{0}$ a differential operator
mapping a $t$ dependent section of a bundle $E^{p,q}$ into another such
section by the formula: 
\begin{equation}
\hat{\partial}_{0}\hat{\nabla}^{p}u^{A}=\bar{\partial}_{0}\hat{\nabla}%
^{p}u^{A}+G_{BC}^{A}\partial _{0}u^{B}\hat{\nabla}^{p}u^{C}
\end{equation}
with \footnote{{\footnotesize Operator on tensors denoted }$\hat{\partial}%
_{0}${\footnotesize \ \ in [CB-Yo]. Note that only }$\partial _{0}u^{A}$%
{\footnotesize \ \ is defined, since u \ is a mapping, not a tensor, and
that }$\bar{\partial}_{0}u^{\prime A}\equiv \partial _{0}u^{\prime A}.$%
{\footnotesize .}} 
\begin{equation}
\bar{\partial}_{0}\equiv \partial _{t}-\mathcal{L}_{\nu }
\end{equation}
where $\mathcal{L}_{\nu }$ denotes the Lie derivative with respect to the
shift $\nu .$ In particular: 
\begin{equation}
\text{\ }\hat{\partial}_{0}u^{\prime A}=\partial _{0}u^{\prime
A}+G_{BC}^{A}\partial _{0}u^{B}u^{\prime C}
\end{equation}
and 
\begin{equation}
\text{\ }\hat{\partial}_{0}G^{AB}=0.
\end{equation}
With these notations the wave map equation reads: 
\begin{equation}
-\hat{\partial}_{0}u^{\prime A}+\hat{\nabla}^{a}(N\partial _{a}u^{A})+N\tau
u^{\prime }=0.
\end{equation}
We will use the following lemma, which can be foreseen, and also checked%
\footnote{{\footnotesize See CB1.}} by direct computation.

\begin{lemma}
The following commutation relations are satisfied: 
\begin{equation}
\hat{\partial}_{0}\partial_{a}u^{A}=\hat{\nabla}_{a}\partial_{0}u^{A},
\end{equation}
\begin{equation}
\hat{\partial}_{0}\hat{\nabla}_{a}\partial_{0}u^{A}-\hat{\nabla}_{a}\hat{%
\partial}_{0}\partial_{0}u^{A}=R_{CB}{}^{A}{}_{D}\partial_{0}u^{C}%
\partial_{a}u^{B}\partial_{0}u^{D},
\end{equation}
\begin{equation}
\hat{\partial}_{0}\hat{\nabla}_{a}\partial_{b}u^{A}-\hat{\nabla}_{a}\hat{%
\partial}_{0}\partial_{b}u^{A}=R_{CB}{}^{A}{}_{D}\partial_{0}u^{C}%
\partial_{a}u^{B}\partial_{b}u^{D}-\partial_{c}u^{A}\hat{\partial}%
_{0}\Gamma_{ab}^{c}.
\end{equation}
\end{lemma}

We recall the identities 
\begin{equation}
\hat{\partial}_{0}g^{ab}=\bar{\partial}_{0}g^{ab}=2Nk^{ab},
\end{equation}
\begin{equation}
\hat{\partial}_{0}\Gamma_{ab}^{c}=\bar{\partial}_{0}\Gamma_{ab}^{c}=\nabla
^{c}(Nk_{ab})-\nabla_{a}(Nk_{b}^{c})-\nabla_{b}(Nk_{a}^{c}).
\end{equation}

\subsubsection{Definition.}

We define the second energy by the following formula

\begin{equation}
E^{(1)}(t)\equiv \int_{\Sigma _{t}}(J_{0}+J_{1})\mu _{g}
\end{equation}
with 
\begin{equation}
J_{1}=\frac{1}{2}\mid \hat{\Delta}_{g}u\mid ^{2}\equiv \frac{1}{2}\{(2(\hat{%
\Delta}_{g}\gamma )^{2}+\frac{1}{2}e^{-4\gamma }(\hat{\Delta}_{g}\omega
)^{2}\}
\end{equation}
\begin{equation}
J_{0}=\frac{1}{2}\mid \hat{\nabla}u^{\prime }\mid _{g}^{2}\equiv \frac{1}{2}%
\{2|\hat{\nabla}\gamma ^{\prime }|_{g}^{2}+\frac{1}{2}e^{-4\gamma }|\hat{%
\nabla}\omega ^{\prime }|_{g}^{2}\}.
\end{equation}

\subsubsection{Estimate.}

We postpone the computation and estimate of the derivative of $E^{(1)}(t)$
until after the estimates of $h,\lambda,N$ and $\nu.$ We set: 
\begin{equation}
E(t)\equiv\varepsilon^{2},\text{ \ \ }E^{(1)}(t)\equiv\tau^{2}\varepsilon
_{1}^{2}
\end{equation}

\subsection{Norms.}

We suppose chosen a smooth cross section $Q\rightarrow\psi(Q)$ of $M_{-1}$
over the Teichmuller space $T_{eich}$, together with a $C^{1}$ curve $%
t\rightarrow Q(t).$ We are then given by lift to $M_{-1}$ a regular metric $%
\sigma_{t}$ for $t\in\lbrack t_{0},T]$, with scalar curvature -1.

\begin{definition}
Hypothesis H$_{\sigma}:$ the curve is contained in a compact subset of $%
T_{eich}$.
\end{definition}

Under the hypothesis H$_{\sigma}$ the metric $\sigma_{t}$ is uniformly
equivalent to the metric $\sigma_{0}\equiv\sigma_{t_{0}}.$ A $t-$ dependent
Sobolev constant on $(\Sigma,\sigma_{t})$ is uniformly equivalent to a
number. We denote by $C_{\sigma}$ any such number, which depends only on the
considered compact subset of $T_{eich}.$

The spaces $W_{s}^{p}(\sigma_{t})$ are the usual Sobolev spaces of tensor
fields on the riemannian manifold $(\Sigma,\sigma_{t}).$ By the hypothesis
on $\sigma_{t}$ their norms are uniformly equivalent for $t\in\lbrack
t_{0,},T]$ to the norm in $W_{s}^{p}(\sigma_{0})$ denoted simply $W_{s}^{p}. 
$ We set $W_{s}^{2}=H_{s}.$

We denote now by $|.|$ a pointwise norm in the $\sigma $ and $G$ metrics; $%
\parallel .\parallel $ and $\parallel .\parallel _{p}$ denote $L^{2}$ and $%
L^{p}$ norms in the $\sigma $ metric.

We denote by $\hat{D}$ a covariant derivative relative to the metrics $%
\sigma $ and $G.$

A lower case index $m$ or $M$ denotes respectively the lower or upper bound
of a scalar function on $\Sigma_{t}$. It may depend on $t$.

\begin{lemma}
It holds that:

1. 
\begin{equation}
||Du||^{2}\equiv||Du||_{g}^{2}\leq2\varepsilon^{2},\text{ \ \ }||u^{\prime
}||^{2}\leq
e^{-2\lambda_{m}}||u^{\prime}||_{g}^{2}\leq2e^{-2\lambda_{m}}\varepsilon^{2}.
\end{equation}

2. 
\begin{equation}
||\hat{D}Du||^{2}\leq2e^{2\lambda_{M}}\tau^{2}\varepsilon_{1}^{2}+%
\varepsilon^{2}.
\end{equation}
\end{lemma}

\begin{proof}
1. Results directly from the definitions.

2. By definition 
\begin{equation}
||\hat{D}Du||^{2}=\int_{\Sigma }\hat{D}^{a}D^{b}u.\hat{D}_{a}D_{b}u\mu
_{\sigma }
\end{equation}
\begin{equation}
=\int_{\Sigma }\{\hat{D}^{a}(D^{b}u.\hat{D}_{a}D_{b}u)-D^{b}u.\hat{D}^{a}%
\hat{D}_{a}D_{b}u\}\mu _{\sigma }=-\int_{\Sigma }D^{b}u.\hat{D}^{a}\hat{D}%
_{a}D_{b}u\mu _{\sigma }
\end{equation}
(since $D^{b}u.\hat{\nabla}_{a}D_{b}u$ is an ordinary covariant vector on $%
\Sigma $ its divergence integrates to zero).

The Ricci commutation formula gives that, with $\rho_{ab}=$ $-\frac{1}{2}%
\sigma_{ab}$ the Ricci curvature of the metric $\sigma$: 
\begin{equation}
\hat{D}^{a}\hat{D}_{a}D_{b}u^{C}=\hat{D}^{a}\hat{D}_{b}D_{a}u^{C}=\hat{D}_{b}%
\hat{\Delta}u^{C}+%
\rho_{b}{}^{c}D_{c}u^{C}+D^{a}u^{A}D_{b}u^{B}R_{AB,}{}^{C}{}_{D}D_{a}u^{D}.
\end{equation}
By another integration by parts 4.31 gives then 
\begin{equation*}
\int_{\Sigma}-D^{b}u.\hat{D}^{a}\hat{D}_{a}D_{b}u\mu_{\sigma}=\int_{\Sigma
}\{|\hat{\Delta}u|^{2}-D^{b}u.(%
\rho_{b}^{c}D_{c}u^{C}+D^{a}u^{A}D_{b}u^{B}R_{AB,}{}^{.}{}_{D}D_{a}u^{D})\}%
\mu_{\sigma}
\end{equation*}

On a 2 dimensional manifold the Riemann curvature is given by: 
\begin{equation}
R_{AB,}{}^{C}{}_{D}=\frac{1}{2}R(G)\{\delta_{A}^{C}G_{BD}-%
\delta_{B}^{C}G_{AD}\}
\end{equation}
A straightforward computation gives therefore 
\begin{equation}
D^{b}u.D^{a}u^{A}D_{b}u^{B}R_{AB,}{}^{.}{}_{D}D_{a}u^{D}=
\end{equation}
\begin{equation}
\frac{1}{2}R(G)D^{b}u^{E}D^{a}u^{A}D_{b}u^{B}D_{a}u^{D}%
\{G_{AE}G_{BD}-G_{BE}G_{AD}\}=
\end{equation}
\begin{equation}
\frac{1}{2}R(G)%
\{(D^{b}u.D^{a}u)(D_{b}u.D_{a}u)-(D^{b}u.D_{b}u)(D^{a}u.D_{a}u)\}=
\end{equation}
\begin{equation}
\frac{1}{2}R(G)\{|Du.Du|^{2}-|\text{ }|Du|^{2}|^{2}\}
\end{equation}
In the case of the Poincar\'{e} plane $R(G)=-4,$ hence 
\begin{equation}
-D^{b}u.D^{a}u^{A}D_{b}u^{B}R_{AB,}{}^{.}{}_{D}D_{a}u^{D}=2\{|Du.Du|^{2}-|%
\text{ }|Du|^{2}|^{2}\}\leq0,
\end{equation}
because 
\begin{equation}
|Du.Du|\leq\text{ }|Du|^{2}.
\end{equation}
\end{proof}

\section{First elliptic estimates.}

The equations for $h,$ $\lambda ,$ $N,$ and $\nu $ are elliptic equations on 
$(\Sigma _{t},\sigma _{t}),$ identical with those written in [CB-M1], except
that in the coefficients $Du.\dot{u},$ $|Du|^{2},$ $|\dot{u}|^{2}$ which
appear in these equations $u$ is now a wave map and not a scalar function.
The estimates obtained in [CB-M1] in terms of $\varepsilon $ and $%
\varepsilon _{1}$ will be valid if the new coefficients satisfy the same
estimates in terms of our new $\varepsilon $ and $\varepsilon _{1}.$

\subsection{Basic bounds on $N$ and $\protect\lambda.$}

The generalized maximum principle\footnote{{\footnotesize The coefficients
in these equations belong to the same functional spaces as in [CB-M1], as
will be proved in the next subsection which will also estimate them.}}
applied to the equations 2.24 and 2.23\ satisfied respectively by $N$ and $%
\lambda$ shows that, with our choice of the time parameter 
\begin{equation}
t=-\tau^{-1},
\end{equation}
it holds that 
\begin{equation}
0\leq N_{m}\leq N\leq N_{M}\leq2,
\end{equation}
\begin{equation}
e^{-2\lambda_{M}}\leq e^{-2\lambda}\leq e^{-2\lambda_{m}}\leq\frac{1}{2}%
\tau^{2}.
\end{equation}

\begin{definition}
We say that the hypothesis\footnote{{\footnotesize This hypothesis replaces
the hypothesis H}$_{c}${\footnotesize \ \ made on v in [CB-M1].}} H$%
_{\lambda }$ is satisfied if there exists a number $c_{\lambda}>1,$
independent of $t,$ such that 
\begin{equation}
\frac{1}{\sqrt{2}}e^{\lambda_{M}}|\tau|\leq c_{\lambda}.
\end{equation}
and we denote by $C_{\lambda\text{ }}$ any positive continuous function of $%
c_{\lambda}\in R^{+}.$
\end{definition}

\subsection{$L^{2}$ estimates of $||$ $|Du|^{2}||$ and \TEXTsymbol{\vert}%
\TEXTsymbol{\vert}$\ |\dot{u}|^{2}||$.}

Under the hypothesis H$_{\sigma}$ and H$_{\lambda}$ there exist numbers $%
C_{\sigma},$ $C_{\lambda}$ such that $u$ satisfies the same inequalities as
in the polarized case, that is:

\begin{lemma}
\begin{equation}
||\text{ }|Du|^{2}||\leq C_{\sigma}C_{\lambda}\{\varepsilon^{2}+\varepsilon
\varepsilon_{1}\}.
\end{equation}
\begin{equation}
\Vert\mid u^{\prime}\mid^{2}\Vert\leq
C_{\sigma}C_{\lambda}\tau^{2}\varepsilon(\varepsilon+\varepsilon_{1}).
\end{equation}
\begin{equation}
\Vert\mid\dot{u}\mid^{2}\Vert\leq
C_{\sigma}C_{\lambda}\tau^{-2}\varepsilon(\varepsilon+\varepsilon_{1}).
\end{equation}
\end{lemma}

\begin{proof}
\ A Sobolev embedding theorem applied to the scalar function $|u^{\prime
}|^{2}$ gives that:\footnote{{\footnotesize See C.B2, here case n=2.}} 
\begin{equation*}
\Vert \mid u^{\prime }\mid ^{2}\Vert \leq C_{\sigma }(\Vert |u^{\prime
}|^{2}\Vert _{1}+\Vert D|u^{\prime }|^{2}\Vert _{1}).
\end{equation*}
It holds that 
\begin{equation*}
D|u^{\prime }|^{2}\equiv 2u^{^{\prime }}.\hat{D}u^{\prime }
\end{equation*}
therefore, since $\hat{D}u^{\prime }\equiv \hat{\nabla}u^{\prime }$, 
\begin{equation*}
\Vert D|u^{\prime }|^{2}\Vert _{1}\leq 2||u^{\prime }||\text{ }||\hat{D}%
u^{\prime }||\leq 2e^{-\lambda _{m}}||u^{\prime }||_{g}\text{ }||\hat{\nabla}%
u^{\prime }||_{g}
\end{equation*}
Hence, since $\Vert |u^{\prime }|^{2}\Vert _{1}\equiv ||u^{\prime }||^{2},$
and using the bound 5.3 of $\lambda $%
\begin{equation*}
\Vert \mid u^{\prime }\mid ^{2}\Vert \leq C_{\sigma }e^{-\lambda
_{m}}||u^{\prime }||_{g}(e^{-\lambda _{m}}||u^{\prime }||_{g}+||\hat{D}%
u^{\prime }||_{g})\leq C_{\sigma }|\tau |^{2}\varepsilon (\varepsilon
+\varepsilon _{1})
\end{equation*}
By the definition of $\dot{u}$ it holds that. 
\begin{equation*}
\parallel |\dot{u}|^{2}\parallel \leq e^{4\lambda _{M}}\parallel |u^{\prime
}|^{2}\parallel ,
\end{equation*}
hence 
\begin{equation*}
\Vert \mid \dot{u}\mid ^{2}\Vert \leq C_{\sigma }e^{4\lambda _{M}}\tau
^{2}\varepsilon (\varepsilon +e^{\lambda _{M}}|\tau |\varepsilon _{1})\leq
C_{\sigma }C_{\lambda }\tau ^{-2}\varepsilon (\varepsilon +\varepsilon _{1}).
\end{equation*}
On the other hand, since 
\begin{equation}
D|Du|^{2}\equiv 2Du.\hat{D}Du,
\end{equation}
it holds that, using again the Sobolev embedding theorem 
\begin{equation*}
\Vert \mid Du\mid ^{2}\Vert \leq C_{\sigma }\Vert Du\Vert (\Vert Du\Vert
+\Vert \hat{D}Du\Vert )
\end{equation*}
which gives using lemma 4.2 
\begin{equation*}
\Vert \mid Du\mid ^{2}\Vert \leq C_{\sigma }\varepsilon (\varepsilon
+e^{\lambda _{M}}|\tau |\varepsilon _{1})
\end{equation*}
hence under the hypothesis H$_{\lambda }:$%
\begin{equation*}
\Vert \mid Du\mid ^{2}\Vert \leq C_{\sigma }C_{\lambda }\varepsilon
(\varepsilon +\varepsilon _{1}).
\end{equation*}
\end{proof}

\begin{lemma}
\begin{equation}
||\text{ }|Du|^{2}||_{g}\leq C_{\sigma}C_{\lambda}|\tau|\varepsilon
\{\varepsilon+\varepsilon_{1}\}.
\end{equation}
\begin{equation}
\Vert\mid u^{\prime}\mid^{2}\Vert_{g}\leq C_{\sigma}C_{\lambda}|\tau
|\varepsilon(\varepsilon+\varepsilon_{1}).
\end{equation}
\end{lemma}

\begin{proof}
The inequalities 5.5 and 5.6 imply that 
\begin{equation*}
\Vert\mid
u^{\prime}\mid^{2}\Vert_{g}\equiv(\int_{\Sigma_{t}}|u^{\prime}|^{4}\mu_{g})^{%
\frac{1}{2}}\leq e^{\lambda_{M}}\Vert\mid u^{\prime}\mid ^{2}\Vert\leq
C_{\lambda}C_{\sigma}|\tau|\varepsilon(\varepsilon +\varepsilon_{1})
\end{equation*}
and, using the lower bound of $\lambda$, 
\begin{equation*}
\Vert\mid Du\mid^{2}\Vert_{g}\equiv(\int_{\Sigma_{t}}|Du|_{g}^{4}\mu _{g})^{%
\frac{1}{2}}\leq e^{-\lambda_{m}}\Vert\mid Du\mid^{2}\Vert\leq
C_{\sigma}C_{\lambda}|\tau|\varepsilon(\varepsilon+\varepsilon_{1}).
\end{equation*}
\end{proof}

\subsection{Estimate of $h$ in $H_{1}.$}

\ We have defined the auxiliary unknown $h$ by 
\begin{equation*}
h_{ab}\equiv k_{ab}-\frac{1}{2}g_{ab}\tau
\end{equation*}

\subsubsection{Estimate of $||h||.$}

The $L^{2}$ norm of $h$ on $(\Sigma,\sigma)$ is bounded in terms of the
first energy and an upper bound $\lambda_{M}$ of the conformal factor since
we have 
\begin{equation*}
\parallel
h\parallel^{2}=\int_{\Sigma_{t}}\sigma^{ac}\sigma^{bd}h_{ab}h_{cd}\mu_{%
\sigma}\leq e^{2\lambda_{M}}\parallel
h\parallel_{L^{2}(g)}^{2}\leq2e^{2\lambda_{M}}\varepsilon^{2}.
\end{equation*}

\subsubsection{Estimate of $\parallel Dh\parallel.$}

The tensor $h$ satisfies the equations 
\begin{equation*}
D_{a}h_{b}^{a}=L_{b}\equiv -\partial _{a}u.\overset{.}{u}
\end{equation*}
It is the sum of a TT (transverse, traceless) tensor $h_{TT}\equiv q$ and a
conformal Lie derivative $r$: 
\begin{equation*}
h\equiv q+r
\end{equation*}
It results from elliptic theory that on each $\Sigma _{t}$ the tensor $r$
satisfies the estimate 
\begin{equation*}
\parallel r\parallel _{H_{1}}\leq C_{\sigma }\parallel Du.\overset{.}{u}%
\parallel \leq C_{\sigma }\parallel |Du|^{2}\parallel ^{\frac{1}{2}%
}\parallel |\overset{.}{u}|^{2}\parallel ^{\frac{1}{2}}.
\end{equation*}
It results from the inequalities of the lemma 5.2 that 
\begin{equation*}
\parallel r\parallel _{H_{1}}\leq C_{\sigma }C_{\lambda }e^{2\lambda
_{M}}|\tau |\varepsilon \{(\varepsilon +\varepsilon _{1})(\varepsilon
+\varepsilon _{1}e^{\lambda _{M}}|\tau |)\}^{\frac{1}{2}}
\end{equation*}

We recall that for the transverse part $h_{TT}=q$ it holds that 
\begin{equation*}
\parallel Dq\parallel=\parallel q\parallel\leq\parallel h\parallel+\parallel
r\parallel
\end{equation*}
therefore 
\begin{equation*}
\parallel Dh\parallel\leq e^{\lambda_{M}}\varepsilon\{\sqrt{2}+C_{\sigma
}e^{\lambda_{M}}|\tau|(\varepsilon+\varepsilon_{1}e^{\lambda_{M}}|\tau|)^{%
\frac{1}{2}}(\varepsilon+\varepsilon_{1})^{\frac{1}{2}}\}.
\end{equation*}

\begin{definition}
We say that the hypothesis H$_{E}$ is satisfied if there exists \ a positive
number $c_{E}$ such that $\varepsilon+\varepsilon_{1}\leq c_{E}.$ We denote
by $C_{E}$ any continuous and positive function of $c_{E}\in R^{+}.$
\end{definition}

If the hypothesis H$_{\sigma},$ H$_{\lambda}$ and H$_{E}$ are satisfied, $h $
verifies the inequality: 
\begin{equation*}
\parallel Dh\parallel\leq C_{\lambda}|\tau|^{-1}\varepsilon(1+C_{\sigma
}C_{\lambda}C_{E}).
\end{equation*}

\subsection{Estimates for the conformal factor $\protect\lambda$.}

The conformal factor $\lambda$ satisfies on each $\Sigma_{t}$ the equation 
\begin{equation*}
\Delta\lambda=f(\lambda)\equiv p_{1}e^{2\lambda}-p_{2}e^{-2\lambda}+p_{3}
\end{equation*}
where the coefficients $p_{i}$ are given by 
\begin{equation*}
p_{1}=\frac{1}{4}\tau^{2},p_{2}=\frac{1}{2}(\mid h\mid^{2}+\mid\dot{u}\mid
^{2}),p_{3}=-\frac{1}{2}(1+\mid Du\mid^{2})
\end{equation*}
The equation admits the subsolution $\lambda_{-}$ given by 
\begin{equation*}
e^{-2\lambda_{-}}=\frac{1}{2}\tau^{2}
\end{equation*}
and it holds that 
\begin{equation*}
\lambda_{-}\leq\lambda_{m}\leq\lambda\leq\lambda_{M}\leq\lambda_{+}
\end{equation*}
with $\lambda_{+}$ a supersolution, for example: 
\begin{equation*}
\text{ }\lambda_{+}=\theta+v-minv
\end{equation*}
where $v$ is the solution with mean value zero on $\Sigma_{t}$ of the linear
equation 
\begin{equation*}
\Delta v=f(\theta)\equiv p_{1}e^{2\theta}-p_{2}e^{-2\theta}+p_{3}
\end{equation*}
with\footnote{{\footnotesize We have renamed }$\theta${\footnotesize \ the
function called }$\omega${\footnotesize \ in [CB-M1].}} $e^{2\theta}$ a $t- $
dependent number, positive solution of the equation 
\begin{equation*}
\bar{p}_{1}e^{4\theta}+\bar{p}_{3}e^{2\theta}-\bar{p}_{2}=0,
\end{equation*}
where $\bar{f}$ denotes the mean value on $(\Sigma,\sigma)$ of a function $%
f: $%
\begin{equation*}
\bar{f}\equiv\frac{1}{V_{\sigma}}\int_{\Sigma}f\mu_{\sigma},\text{ \ \ }%
V_{\sigma}\equiv\int_{\Sigma}\mu_{\sigma}=-4\pi\chi.
\end{equation*}
Using the expressions of $\bar{p}_{2},\bar{p}_{3}$ and $\bar{p}_{1}=\frac
{1}{4}\tau^{2},$ together with 
\begin{equation*}
\parallel\overset{.}{u}\parallel^{2}\leq e^{2\lambda_{M}}\parallel u^{\prime
}\parallel_{g}^{2},\text{ and }\parallel h\parallel^{2}\leq
e^{2\lambda_{M}}\parallel h\parallel_{g}^{2}
\end{equation*}
and the expression of $\varepsilon^{2}\equiv E(t)$ we have found in [CB-M1],
section 8.1 (recall that $V_{\sigma}=-4\pi\chi,$ a constant) that (we have
renamed $\theta$ the $t-$ dependent number $\omega$ of CB-M1):

\begin{equation}
0\leq\frac{1}{2}\tau^{2}e^{2\theta}-1\leq V_{\sigma}^{-1}(1+\frac{\tau^{2}}{2%
}e^{2\lambda_{M}})\varepsilon^{2}\leq C_{\lambda}\varepsilon^{2}
\end{equation}

\begin{lemma}
The following inequalities hold

1. 
\begin{equation}
0\leq\lambda_{M}-\theta\leq2\parallel v\parallel_{L^{\infty}}.
\end{equation}

2. 
\begin{equation}
1\leq\frac{1}{2}\tau^{2}e^{2\lambda_{M}}\leq1+C_{\lambda}\varepsilon
^{2}+C_{\lambda}C_{E}\parallel v\parallel_{L^{\infty}}e^{4\parallel
v\parallel_{L^{\infty}}}
\end{equation}
\end{lemma}

\begin{proof}
1.\ It holds that 
\begin{equation}
\lambda_{M}\leq\sup\lambda_{+}=\theta+maxv-minv,
\end{equation}
from which results 5.12.

2.\ The inequality 5.12 implies by elementary calculus 
\begin{equation}
e^{2(\lambda_{M}-\theta)}\leq1+4||v||_{L^{\infty}}e^{4||v||_{L^{\infty}}}
\end{equation}
.The inequalities 5.11 and 5.15 imply that 
\begin{equation}
\frac{1}{2}\tau^{2}e^{2\lambda_{M}}\leq(1+C_{\lambda}\varepsilon
^{2})(1+4\parallel v\parallel_{L^{\infty}}e^{4\parallel
v\parallel_{L^{\infty }}}),
\end{equation}
from which the inequality 5.13 follows.
\end{proof}

\subsubsection{Estimate of v.}

The equation satisfied by $v$ implies 
\begin{equation*}
\int_{\Sigma}|Dv|^{2}\mu_{\sigma}=-\int_{\Sigma}f(\theta)v\mu_{\sigma}
\end{equation*}
hence 
\begin{equation*}
\parallel Dv\parallel^{2}\leq\parallel f(\theta)\parallel\parallel v\parallel
\end{equation*}
but the Poincar\'{e} inequality applied to the function $v$ which has mean
value 0 on $\Sigma$ gives 
\begin{equation*}
\parallel v\parallel^{2}\leq\lbrack\Lambda_{\sigma}]^{-1}\parallel
Dv\parallel^{2}
\end{equation*}
where $\Lambda_{\sigma}$ is the first (positive) eigenvalue of - $%
\Delta_{\sigma}$ for functions on $\Sigma_{t}$ with mean value zero.
Therefore on each $\Sigma_{t}$%
\begin{equation*}
\parallel Dv\parallel\leq\lbrack\Lambda_{\sigma}]^{-1/2}\parallel
f(\theta)\parallel
\end{equation*}
We use Ricci identity and $R(\sigma)=-1$ to obtain that 
\begin{equation*}
\parallel\Delta_{\sigma}v\parallel^{2}=\parallel D^{2}v\parallel^{2}-\frac
{1}{2}\parallel Dv\parallel^{2}
\end{equation*}
The equation satisfied by $v$ implies then, as in [1], 
\begin{equation*}
\Vert D^{2}v\Vert^{2}=\Vert f(\theta)\Vert^{2}+\frac{1}{2}\Vert Dv\Vert^{2}
\end{equation*}
Assembling these various inequalities gives that: 
\begin{equation*}
\parallel v\parallel_{H_{2}}\leq\lbrack1+3/(2\Lambda_{\sigma})+1/\Lambda
_{\sigma}^{2}]^{1/2}\parallel f(\theta)\parallel
\end{equation*}
The Sobolev inequality 
\begin{equation*}
\parallel v\parallel_{L^{\infty}}\leq C_{\sigma}\parallel v\parallel_{H_{2}}
\end{equation*}
gives then a bound on the $L^{\infty}$ norm of $v$ on $\Sigma_{t\text{ }}$in
terms of the $L^{2}$ norm of $f(\theta)$, a Sobolev constant $C_{\sigma} $
and the lowest eigenvalue $\Lambda_{\sigma}$ of $-\Delta_{\sigma},$ which is
itself a number $C_{\sigma}.$

We now estimate the $L^{2}$ norm of $f(\theta).$%
\begin{equation*}
f(\theta)\equiv f\equiv p_{1}e^{2\theta}-p_{2}e^{-2\theta}+p_{3}.
\end{equation*}
By the isoperimetric inequality, and since $\bar{f}=0,$ there exists a
constant $C_{\sigma}$ such that: 
\begin{equation*}
\Vert f\Vert\leq C_{\sigma}\Vert Df\Vert_{1}
\end{equation*}
We want to bound the right hand side in terms of the first and second
energies of the wave map. We have by the definition of $f$ and the
expression of the $p^{\prime}s$ that: 
\begin{equation*}
\Vert Df\Vert_{1}\leq\frac{1}{2}\{\Vert
D|Du|^{2}\Vert_{1}+e^{-2\theta}(\Vert D|h|^{2}\Vert_{1}+\Vert D|\dot{u}%
|^{2}\Vert_{1})\}
\end{equation*}

\begin{lemma}
The following estimate holds under the hypothesis H$_{\lambda}$: 
\begin{equation*}
\frac{1}{2}\parallel D|Du|^{2}\parallel_{1}\leq C_{\lambda}(\varepsilon
^{2}+\varepsilon\varepsilon_{1})
\end{equation*}
\end{lemma}

\begin{proof}
We have: 
\begin{equation*}
D|Du|^{2}=2Du.\hat{D}^{2}u
\end{equation*}
hence 
\begin{equation*}
\Vert D|Du|^{2}\Vert_{1}\leq2\Vert Du\Vert\Vert\hat{D}^{2}u\Vert
\end{equation*}
We have seen that 
\begin{equation}
\Vert Du\Vert=\parallel Du\parallel_{g}\text{, \ \ \ }\Vert\hat{D}%
^{2}u\Vert\leq e^{\lambda_{M}}\Vert\hat{\Delta}_{g}u\Vert_{g}+(1/\sqrt{2}%
)\Vert Du\Vert_{g}
\end{equation}
which implies the given result under the hypothesis H$_{\lambda}$ and the
definitions of $\varepsilon$ and $\varepsilon_{1}.$
\end{proof}

\begin{lemma}
The following estimates hold under the hypothesis H$_{\sigma},$ H$_{E}$ and $%
H_{\lambda}:$

1. 
\begin{equation*}
\frac{1}{2}e^{-2\theta}\parallel D|h|^{2}\parallel_{1}\leq
C_{\sigma}C_{E}C_{\lambda}\varepsilon^{2}
\end{equation*}

2. 
\begin{equation}
\frac{1}{2}e^{-2\theta}\Vert D|\dot{u}|^{2}\Vert_{1}\leq C_{E}C_{\lambda
}C_{\sigma}(\varepsilon^{2}+\varepsilon\varepsilon_{1}).
\end{equation}
\end{lemma}

\begin{proof}
1.\ We have: 
\begin{equation*}
\parallel D|h^{2}|\parallel_{1}\leq2\parallel h\parallel\parallel Dh\parallel
\end{equation*}
Using the inequalities of sections 5.3.1 and 5.3.2 we find that 
\begin{equation*}
\parallel D|h^{2}|\parallel_{1}\leq C_{E}C_{\lambda}\tau^{-2}\varepsilon^{2}
\end{equation*}
The given result follows from the bound 5.11\ of $e^{-2\theta}.$

2.\ The estimate given in [CB-M1], lemma 21, when $u$ is a scalar function
holds when $u$ is a wave map, with the same proof which, as far as $u$ is
concerned, contains only norms. It gives the announced inequality.
\end{proof}

We denote by $C_{E,\lambda,\sigma}$ a number depending only on $%
c_{E},c_{\lambda}$ and the considered compact domain of $T_{eich}.$

\begin{lemma}
There exists a number $C_{E,\lambda,\sigma}$ such that the L$^{\infty}$ norm
of v is bounded by the following inequality 
\begin{equation}
\parallel v\parallel_{\infty}\leq C_{E,\lambda,\sigma}(\varepsilon
^{2}+\varepsilon\varepsilon_{1})
\end{equation}
\end{lemma}

\begin{proof}
Recall that there exists a Sobolev constant $C_{\sigma}$ such that 
\begin{equation*}
\Vert v\Vert_{\infty}\leq C_{\sigma}\{\Vert D|Du|^{2}\Vert_{1}+e^{-2\omega
}(\Vert D|h|^{2}\Vert_{1}+\Vert D|\dot{u}|^{2}\Vert_{1})\}
\end{equation*}
The three terms in the sum have been evaluated in the lemma 5.4.
\end{proof}

\begin{theorem}
1.It holds that: 
\begin{equation}
1\leq\frac{1}{2}\tau^{2}e^{2\lambda_{M}}\leq1+C_{E,\lambda,\sigma}(%
\varepsilon+\varepsilon_{1})^{2},
\end{equation}

2. There exists a number $\eta_{1}>0$ such that the hypothesis H$_{\lambda} $
is satisfied, i.e.: 
\begin{equation}
1\leq\frac{1}{\sqrt{2}}|\tau|e^{\lambda_{M}}\leq c_{\lambda},\text{ \ \ }%
c_{\lambda}>1,
\end{equation}
as soon as 
\begin{equation}
\varepsilon+\varepsilon_{1}\leq\eta_{1}.
\end{equation}
\end{theorem}

\begin{proof}
1. The lemmas 5.3 and 5.6.

2.\ By 5.23 it holds that, with $C_{E,\lambda,\sigma}$ the number of that
inequality, 
\begin{equation}
1\leq\frac{1}{\sqrt{2}}|\tau|e^{\lambda_{M}}<c_{\lambda}\text{ \ \ if \ \ (}%
\varepsilon+\varepsilon_{1})^{2}<\frac{c_{\lambda}^{2}-1}{C_{E,\lambda
,\sigma}}.
\end{equation}
The result follows from a continuity argument.
\end{proof}

\subsubsection{Bound of $\protect\lambda$ in $H_{1}.$}

The following theorem holds, with the same proof as the theorem 23 of
[CB-M1].

\begin{theorem}
The following estimate holds 
\begin{equation}
||D\lambda||_{H_{1}}\leq C_{E,\lambda,\sigma}(\varepsilon^{2}+\varepsilon
\varepsilon_{1}).
\end{equation}
\end{theorem}

\section{Estimates in $W_{s}^{p}.$}

\subsection{Estimates for h in $W_{2}^{p}$.}

The estimates of $h$ in $W_{2}^{p}$, with $1<p<2$ (for definiteness we will
choose $p=\frac{4}{3})$ will be obtained using estimates for the conformal
factor $\lambda$ which have been obtained by using the $H_{1}$ norm of $h.$

\begin{theorem}
Under the H hypotheses there exists a positive number $C_{E,\lambda,\sigma} $
such that the $W_{2}^{p}$ norm of h, choosing to be specific $p=\frac{4}{3},$
is bounded by 
\begin{equation*}
\parallel h\parallel_{W_{2}^{p}}\leq
C_{E,\lambda,\sigma}|\tau|^{-1}(\varepsilon+\varepsilon_{1})
\end{equation*}
\end{theorem}

\begin{corollary}
It holds that 
\begin{equation*}
|\tau|\parallel h\parallel_{\infty}\leq C_{E,\lambda,\sigma}(\varepsilon
+\varepsilon_{1})
\end{equation*}
and that 
\begin{equation*}
\parallel h\parallel_{L^{\infty}(g)}\leq C_{E,\lambda,\sigma}|\tau
|(\varepsilon+\varepsilon_{1}).
\end{equation*}
\end{corollary}

\begin{proof}
The inequalities satisfied by $||h||_{W_{2}^{p}}$ in [CB-M1], with $p=\frac
{4}{3},$ are still valid when $u$ is a wave map, with the same proof, in
particular because $\parallel Du.\overset{.}{u}\parallel_{\frac{4}{3}}$ and $%
\parallel D(Du.\overset{.}{u})\parallel_{\frac{4}{3}}$ satisfy estimates of
the same type as in [CB-M1]; indeed, using section 5.2: 
\begin{equation*}
\parallel Du.\overset{.}{u}\parallel_{\frac{4}{3}}\leq\parallel Du\parallel
\parallel\overset{.}{u}\parallel_{4}\leq C_{E,\lambda,\sigma}|\tau
|^{-1}\varepsilon^{\frac{3}{2}}(\varepsilon+\varepsilon_{1})^{\frac{1}{2}}.
\end{equation*}
On the other hand a straightforward calculation gives 
\begin{equation}
D(Du.\overset{.}{u})\equiv\hat{D}(Du).\dot{u}+Du.\hat{D}\dot{u}
\end{equation}
hence 
\begin{equation*}
\parallel D(Du.\overset{.}{u})\parallel_{\frac{4}{3}}\leq\parallel\hat{D}%
^{2}u\parallel\parallel\overset{.}{u}\parallel_{4}+\parallel|Du|\parallel
_{4}\parallel\hat{D}\overset{.}{u}\parallel
\end{equation*}
which gives, using previous estimates 
\begin{equation*}
\parallel D(Du.\overset{.}{u})\parallel_{\frac{4}{3}}\leq
C_{\lambda}C_{\sigma}e^{\lambda_{M}}\{\varepsilon^{\frac{1}{2}}(\varepsilon
+\varepsilon_{1})^{\frac{3}{2}}+\varepsilon^{\frac{3}{2}}(\varepsilon
+\varepsilon_{1})^{\frac{1}{2}}\}
\end{equation*}
The result of the theorem follows from the bound of $\varepsilon$ by $%
\varepsilon+\varepsilon_{1}$.

The corollary is a consequence of the Sobolev embedding theorem, 
\begin{equation*}
\parallel h\parallel_{\infty}\leq C_{\sigma}\parallel h\parallel_{W_{2}^{p}}%
\text{ \ \ \ if \ \ \ }p>1,
\end{equation*}
and the estimate . 
\begin{equation*}
\parallel
h\parallel_{L^{\infty}(g)}=Sup_{\Sigma}\{g^{ac}g^{bd}h_{ab}h_{cd}\}^{\frac{1%
}{2}}\leq e^{-2\lambda_{m}}\parallel h\parallel_{\infty}\leq\frac{1}{2}%
\tau^{2}\parallel h\parallel_{\infty}
\end{equation*}
\end{proof}

\subsection{W$_{3}^{p}$ estimates for N.}

\subsubsection{$H_{2}$ estimates of N.}

\begin{theorem}
There exists a number $C_{E,\lambda,\sigma}$ such that the $H_{2}$norm of $N$
satisfies the inequality 
\begin{equation*}
\parallel2-N\parallel_{H_{2}}\leq C_{E,\lambda,\sigma}(\varepsilon
^{2}+\varepsilon\varepsilon_{1})
\end{equation*}
\end{theorem}

\begin{corollary}
a. It holds that: 
\begin{equation}
\parallel2-N\parallel_{L^{\infty}}\leq C_{E,\lambda,\sigma}(\varepsilon
^{2}+\varepsilon\varepsilon_{1})
\end{equation}
b. There exists $\eta_{2}>0$ such that 
\begin{equation}
\varepsilon+\varepsilon_{1}\leq\eta_{2}
\end{equation}
implies the existence of a positive number $N_{m}$\ (independent of $t)$
such that 
\begin{equation*}
N\geq N_{m}>0.
\end{equation*}
\end{corollary}

\begin{proof}
We write, as in [CB-M1] the equation satisfied by $N$ in the form 
\begin{equation*}
\Delta(2-N)-(2-N)=\beta
\end{equation*}
with, having chosen the parameter $t$ such that $\partial_{t}\tau=\tau^{2},$%
\begin{equation*}
\beta\equiv(2-N)(e^{2\lambda}\frac{1}{2}\tau^{2}-1)-N(e^{2\lambda}\mid
u^{\prime}\mid^{2}+e^{-2\lambda}\mid h\mid^{2})
\end{equation*}
The standard elliptic estimate applied to the form given to the lapse
equation gives 
\begin{equation*}
\parallel2-N\parallel_{H_{2}}\leq C_{\sigma}\parallel\beta\parallel
\end{equation*}
Since $0<N\leq2$ and $e^{-2\lambda}\leq\frac{1}{2}\tau^{2}$ it holds that 
\begin{equation}
\parallel\beta\parallel\leq2(\frac{1}{2}e^{2\lambda_{M}}\tau^{2}-1)V_{\sigma
}^{1/2}+2(e^{2\lambda_{M}}\parallel|u^{\prime}|^{2}\parallel+\frac{1}{2}%
\tau^{2}\parallel|h|^{2}\parallel)
\end{equation}
The $L^{4}$ norms of $h$ and $u^{\prime}$ as well as $\frac{1}{2}%
e^{2\lambda_{M}}\tau^{2}-1$ have been estimated in the section on the
conformal factor estimate. We deduce from these estimates the bound 
\begin{equation*}
\parallel\beta\parallel\leq
C_{E,\lambda,\sigma}(\varepsilon^{2}+\varepsilon\varepsilon_{1}).
\end{equation*}
which gives the result of the theorem.

The corollary a. is a consequence of the Sobolev embedding theorem, b. is a
consequence of a.
\end{proof}

\subsubsection{$L^{\infty}$ estimate of $DN.$}

\begin{theorem}
Under the hypotheses H there exist numbers $C_{E},C_{\lambda}$ and $%
C_{\sigma }$ such that if $1<p<2$, for instance $p=\frac{4}{3}$ 
\begin{equation}
\parallel2-N\parallel_{W_{3}^{p}}\leq
C_{\lambda}C_{\sigma}C_{E}(\varepsilon^{2}+\varepsilon\varepsilon_{1}).
\end{equation}

\begin{corollary}
The gradient of $N$ satisfies the inequality: 
\begin{equation}
\parallel DN\parallel_{L^{\infty}(g)}|\leq
C_{\lambda}C_{\sigma}C_{E}|\tau|(\varepsilon^{2}+\varepsilon\varepsilon_{1})
\end{equation}
\end{corollary}
\end{theorem}

\begin{proof}
The proof, essentially the same as in [CB-M1] rests on the $W_{1}^{p}$
estimate of $\beta ,$ since applying the standard elliptic estimate gives 
\begin{equation}
\parallel 2-N\parallel _{W_{3}^{p}}\leq C_{\sigma }\parallel \beta \parallel
_{W_{1}^{p}}
\end{equation}
The estimate of 
\begin{equation*}
\parallel \beta \parallel _{p}\leq V_{\sigma }^{\frac{1}{p}-\frac{1}{2}%
}\parallel \beta \parallel
\end{equation*}
is the same as in [CB-M1] theorem 28. In the estimate $\parallel D\beta
\parallel _{p}$the difference could be only in the estimate of the term $%
\parallel D|u^{\prime }|^{2}\parallel _{p}$. We have here 
\begin{equation}
D|u^{\prime }|^{2}\equiv 2u^{\prime }.\hat{D}u^{\prime }
\end{equation}
hence, with $p=\frac{4}{3}$%
\begin{equation*}
\parallel D|u^{\prime }|^{2}\parallel _{p}\leq 2\parallel u^{\prime
}\parallel _{4}\parallel Du^{\prime }\parallel +2e^{4(\gamma _{M}-\gamma
_{m})}||D\gamma ||_{4}||e^{-2\gamma }\omega ^{\prime }||_{4}^{2}
\end{equation*}
which leads to the same estimate as in [CB-M1]: 
\begin{equation*}
||\beta ||_{W_{1}^{p}}\leq C_{E,\lambda ,\sigma }(\varepsilon
^{2}+\varepsilon \varepsilon _{1}).
\end{equation*}

The corollary is a consequence of the Sobolev embedding theorem and the
relation between $\sigma$ and $g$ norms: 
\begin{equation*}
\parallel DN\parallel_{L^{\infty}(g)}\leq e^{-\lambda_{m}}\parallel
DN\parallel_{\infty}\leq e^{-\lambda_{m}}C_{\sigma}\parallel DN\parallel
_{W_{2}^{p}}\leq C_{E,\lambda,\sigma}|\tau|(\varepsilon^{2}+\varepsilon
\varepsilon_{1})
\end{equation*}
\end{proof}

\section{$\partial_{t}\protect\sigma$ and shift estimates.}

\subsection{$\partial_{t}\protect\sigma$ estimate.}

We have chosen a section $\psi$ of $M_{-1}$ over Teichmuller space, denoted $%
\sigma\equiv\psi(Q)$ and supposed (Hypothesis H$_{\sigma})$ that $Q$ remains
in a compact subset of $T_{eich}.$

We then have: 
\begin{equation}
\partial _{t}\sigma _{ab}=\frac{\partial \psi _{ab}}{\partial Q^{I}}\frac{%
dQ^{I}}{dt}
\end{equation}
where $\frac{\partial \psi }{\partial Q^{I}}$ is uniformly bounded. Hence it
holds that 
\begin{equation}
|\partial _{t}\sigma |\leq C_{\sigma }|\frac{dQ}{dt}|.
\end{equation}
We recall that $Q$ satisfies the differential equation 
\begin{equation*}
X_{IJ}\frac{dQ^{I}}{dt}+Y_{IJ}P^{I}+Z_{J}=0
\end{equation*}
where $X_{IJ}\equiv \int_{\Sigma _{t}}X_{I}^{ab}X_{J,ab}\mu _{\sigma _{t}}$
is a matrix $X$ with uniformly bounded inverse while $Y_{IJ}$ and $Z_{J}$
admit the following bounds, deduced from the basic estimates of $N,\lambda $
and the $L^{2}$ bound of $r:$ 
\begin{equation*}
\text{\TEXTsymbol{\vert}}Y_{IJ}|\equiv |\int_{\Sigma _{t}}2Ne^{-2\lambda
}X_{I}^{ab}X_{J,ab}\mu _{\sigma _{t}}|\leq C_{\sigma }\tau ^{2}
\end{equation*}
\begin{equation*}
|Z_{J}|\equiv |\int_{\Sigma _{t}}2Ne^{-2\lambda }r_{ab}X_{J}^{ab}\mu
_{\sigma _{t}}|\leq C_{\sigma }\tau ^{2}||r||\leq C_{\sigma }C_{E}|\tau
|\varepsilon (\varepsilon +\varepsilon _{1}).
\end{equation*}
On the other hand we recall that 
\begin{equation}
q_{ab}\equiv h_{ab}^{TT}\equiv X_{I,ab}P^{I}
\end{equation}
hence 
\begin{equation}
P^{I}\equiv (X^{-1})^{IJ}\int_{\Sigma _{t}}X_{J}^{ab}q_{ab}\mu _{\sigma }.
\end{equation}
Therefore: 
\begin{equation}
|P^{I}|\leq C_{\sigma }||q||\leq C_{\sigma }(||h||+||r||)\leq C_{\sigma
}C_{E}|\tau |^{-1}(\varepsilon +\varepsilon _{1})
\end{equation}
hence 
\begin{equation}
|Y_{IJ}P^{J}|\leq C_{\sigma }C_{E}|\tau |(\varepsilon +\varepsilon _{1}).
\end{equation}
We have obtained inequalities of the following type: 
\begin{equation}
|\frac{dQ}{dt}|\leq C_{\sigma }C_{E}|\tau |(\varepsilon +\varepsilon _{1}),%
\text{ \ \ }|\partial _{t}\sigma |\leq C_{\sigma }C_{E}|\tau |(\varepsilon
+\varepsilon _{1}).
\end{equation}
The derivatives $D^{k}\partial _{t}\sigma $ satisfy inequalities of the same
type.

\subsection{Shift estimate.}

The equation \ to be satisfied by the shift $\nu$ reads, with $n_{a}$ the
covariant components of the vector $\nu$ in the metric $\sigma,$ i.e. $%
n_{a}\equiv\sigma_{ab}\nu^{b}\equiv e^{-2\lambda}g_{ab}\nu^{b}$:

\begin{equation}
(L_{\sigma }n)_{ab}\equiv D_{a}n_{b}+D_{b}n_{a}-\sigma _{ab}D_{c}n^{c}=f_{ab}
\end{equation}
\begin{equation*}
f_{ab}\equiv 2Ne^{-2\lambda }h_{ab}+\partial _{t}\sigma _{ab}-\frac{1}{2}%
\sigma _{ab}\sigma ^{cd}\partial _{t}\sigma _{cd}
\end{equation*}
The elliptic theory for this first order system gives the estimate 
\begin{equation}
||n||_{W_{3}^{p}}\equiv ||\nu ||_{W_{3}^{p}}\leq C_{\sigma }||f||_{W_{2}^{p}}
\end{equation}
with, if $p>1,$ using the bound 5.3 of $e^{-2\lambda },$%
\begin{equation}
||f||_{W_{2}^{p}\text{ }}\leq C_{\sigma }C_{E}\{\tau
^{2}||N||_{W_{2}^{p}}||\lambda ||_{W_{2}^{p}}||h||_{W_{2}^{p}}+||\partial
_{t}\sigma ||_{W_{2}^{p}}\}.
\end{equation}
Hence, using previous estimates 
\begin{equation}
||f||_{W_{2}^{p}\text{ }}\leq C_{\sigma }C_{E}|\tau |(\varepsilon
+\varepsilon _{1}).
\end{equation}

\section{Second energy estimate.}

We have defined the energy $E^{(1)}(t)$ of gradient $u$ by the formula 
\begin{equation}
\tau^{2}\varepsilon_{1}^{2}\equiv
E^{(1)}(t)\equiv\int_{\Sigma_{t}}(J_{0}+J_{1})\mu_{g}
\end{equation}
with 
\begin{equation}
J_{1}=\frac{1}{2}\mid\hat{\Delta}_{g}u\mid^{2}\equiv\frac{1}{2}\{2(\hat
{\Delta}_{g}\gamma)^{2}+\frac{1}{2}e^{-4\gamma}(\hat{\Delta}_{g}\omega)^{2}\}
\end{equation}
\begin{equation}
J_{0}=\frac{1}{2}\mid\hat{D}u^{\prime}\mid_{g}^{2}\equiv\frac{1}{2}\{2|\hat
{D}\gamma^{\prime}|_{g}^{2}+\frac{1}{2}e^{-4\gamma}|\hat{D}\omega^{\prime
}|_{g}^{2}\}
\end{equation}

\subsection{Second energy equality.}

We have: 
\begin{equation}
\frac{d}{dt}\int_{\Sigma_{t}}(J_{1}+J_{0})\mu_{g}=\int_{\Sigma_{t}}\{%
\partial_{t}(J_{1}+J_{0})-(N\tau-\nabla_{a}\nu^{a})(J_{1}+J_{0})\}\mu_{g}
\end{equation}
On a compact manifold $\Sigma$, divergences integrate to zero, which leads
to the following formula where the shift does not appear explicitly: 
\begin{equation}
\frac{d}{dt}\int_{\Sigma_{t}}(J_{1}+J_{0})\mu_{g}=\int_{\Sigma_{t}}\{%
\partial_{0}(J_{1}+J_{0})-N\tau(J_{1}+J_{0})\}\mu_{g}
\end{equation}
with, since $\hat{\partial}_{0}G_{AB}=0,$ 
\begin{equation}
\partial_{0}J_{1}=\hat{\partial}_{0}\hat{\Delta}_{g}u.\hat{\Delta}_{g}u
\end{equation}

We deduce from the commutation relation of the lemma 4.1 that 
\begin{equation}
\hat{\partial}_{0}\hat{\Delta}_{g}u^{A}=g^{ab}(\hat{\nabla}_{a}\hat{\partial 
}_{0}\partial_{b}u^{A}-\partial_{c}u^{A}\hat{\partial}_{0}\Gamma_{ab}^{c})+%
\hat{\partial}_{0}g^{ab}\hat{\nabla}_{a}\partial_{b}u^{A}+\hat{F}_{1}^{A}
\end{equation}
with 
\begin{equation}
\hat{F}_{1}^{A}\equiv g^{ab}R_{CB}{}^{A}{}_{D}\partial_{0}u^{C}\partial
_{a}u^{B}\partial_{b}u^{D}.
\end{equation}
Hence, using the identities 4.23: 
\begin{equation}
\hat{\partial}_{0}\hat{\Delta}_{g}u^{A}=g^{ab}\hat{\nabla}_{a}\hat{\partial }%
_{0}\partial_{b}u^{A}+N\tau\hat{\Delta}_{g}u^{A}+F_{1}^{A}+\hat{F}_{1}^{A},
\end{equation}
with 
\begin{equation}
F_{1}^{A}\equiv2\partial_{c}u^{A}(h_{g}^{ac}\partial_{a}N+N%
\nabla_{a}k^{ac})+2Nh_{g}^{ab}\hat{\nabla}_{a}\partial_{b}u^{A}.
\end{equation}
We have therefore, using Stokes formula, and $\hat{\partial}_{0}\partial
_{b}u\equiv\hat{\nabla}_{b}(Nu^{\prime})$ 
\begin{equation}
\int_{\Sigma_{t}}\partial_{0}J_{1}\mu_{g}=\int_{\Sigma_{t}}\{-N\hat{\nabla }%
_{a}u^{\prime}.\hat{\nabla}^{a}\hat{\Delta}_{g}u+2N\tau J_{1}-\partial
_{a}Nu^{\prime}.\hat{\nabla}^{a}\hat{\Delta}_{g}u+(F_{1}+\hat{F}_{1}).\hat{%
\Delta}_{g}u\}\mu_{g}
\end{equation}

On the other hand 
\begin{equation*}
\partial _{0}J_{0}=g^{ab}\hat{\partial}_{0}\hat{\nabla}_{a}u^{\prime }.\hat{%
\nabla}_{b}u^{\prime }+N(h_{g}^{ab}+\frac{1}{2}g^{ab}\tau )\partial
_{a}u^{\prime }.\partial _{b}u^{\prime }
\end{equation*}
where we have used the identity, $h_{g}^{ab}$ denoting the contravariant
components of $h_{ab}$ computed with the metric $g,$ 
\begin{equation*}
\hat{\partial}_{0}g^{ab}=2Nk^{ab}\equiv 2Nh_{g}^{ab}+Ng^{ab}\tau .
\end{equation*}
The commutation relation\ 4.20 gives that: 
\begin{equation}
g^{ab}\hat{\partial}_{0}\hat{\nabla}_{a}u^{\prime }.\hat{\nabla}%
_{b}u^{\prime }\equiv \hat{\partial}_{0}\hat{\nabla}_{a}u^{\prime }.\hat{%
\nabla}^{a}u^{\prime }=\hat{\nabla}_{a}\hat{\partial}_{0}u^{\prime }.\hat{%
\nabla}^{a}u^{\prime }+\hat{F}_{0},
\end{equation}
with 
\begin{equation}
\hat{F}_{0}\equiv R_{AB,CD}u^{\prime D}\partial _{0}u^{A}\partial _{a}u^{B}%
\hat{\nabla}^{a}u^{\prime C}
\end{equation}
Therefore, using the wave map equation 
\begin{equation}
-\hat{\partial}_{0}u^{\prime }+N\hat{\Delta}_{g}u+\partial ^{a}N\partial
_{a}u+N\tau u^{\prime }=0.
\end{equation}
we find that 
\begin{equation}
\hat{\partial}_{0}\hat{\nabla}_{a}u^{\prime }.\hat{\nabla}^{a}u^{\prime }=%
\hat{\nabla}_{a}[N\hat{\Delta}_{g}u+\partial ^{c}N\partial _{c}u+N\tau
u^{\prime }].\hat{\nabla}^{a}u^{\prime }+\hat{F}_{0}
\end{equation}
Therefore 
\begin{equation}
\int_{\Sigma _{t}}\partial _{0}J_{0}\mu _{g}=\int_{\Sigma _{t}}\{N\hat{\nabla%
}_{a}\hat{\Delta}_{g}u.\hat{\nabla}^{a}u^{\prime }+3N\tau J_{0}+F_{0}+\hat{F}%
_{0}\}\mu _{g}
\end{equation}
with 
\begin{equation}
F_{0}\equiv \lbrack \partial ^{a}N\hat{\Delta}_{g}u+\hat{\nabla}%
^{a}(\partial ^{c}N\partial _{c}u)].\hat{\nabla}_{a}u^{\prime }+\tau
\partial ^{a}Nu^{\prime }.\hat{\nabla}_{a}u^{\prime }+Nh_{g}^{ab}\hat{\nabla}%
_{a}u^{\prime }.\hat{\nabla}_{b}u^{\prime }
\end{equation}

We see that the third order terms in $u$ disappear from the integral of $%
\partial _{0}(J_{0}+J_{1})$ which reduces to 
\begin{equation}
\int_{\Sigma _{t}}\partial _{0}(J_{0.}+J_{1})\mu _{g}=\int_{\Sigma
_{t}}\{3N\tau J_{0}+2N\tau J_{1}\}\mu _{g}+Z_{1}
\end{equation}
with 
\begin{equation}
Z_{1}\equiv \int_{\Sigma _{t}}\{(F_{1}+\hat{F}_{1}).\hat{\Delta}_{g}u+F_{0}+%
\hat{F}_{0}\}\mu _{g}
\end{equation}
We have obtained 
\begin{equation}
\frac{dE^{(1)}}{dt}=\int_{\Sigma _{t}}N\tau (2J_{0}+J_{1})\mu _{g}+Z_{1}
\end{equation}
which we write 
\begin{equation}
\frac{dE^{(1)}}{dt}-2\tau E^{(1)}=\tau \int_{\Sigma _{t}}NJ_{0}\mu
_{g}+Z_{2}+Z_{1}
\end{equation}
with 
\begin{equation}
Z_{2}\equiv \tau \int_{\Sigma _{t}}(N-2)(2J_{0}+J_{1})\mu _{g}.
\end{equation}

\subsection{Second energy inequality.}

Since $\tau$ is negative (and $N$ positive) the equality 8.21\ implies the
inequality 
\begin{equation}
\frac{dE^{(1)}}{dt}-2\tau E^{(1)}(t)\leq Z_{1}+Z_{2}.
\end{equation}

We now estimate the various terms of $Z_{1},Z_{2},$ called non linear terms
because they are all homogeneous and cubic in $h_{g}^{ab},$ $N-2,$ and the
derivatives of $N$ and $u.$ These estimates are essentially the same as the
ones given in [CB-M1], due to the estimates of the previous section. We
first write, using the estimate of \TEXTsymbol{\vert}\TEXTsymbol{\vert}$%
N-2||_{L^{\infty }(g)}$ and the definition of $\varepsilon _{1}:$%
\begin{equation}
|Z_{2}|\equiv |\tau \int_{\Sigma _{t}}(N-2)(2J_{0}+J_{1})\mu _{g}|\leq
C_{\lambda }C_{\sigma }C_{E}|\tau |^{3}(\varepsilon +\varepsilon _{1})^{4}.
\end{equation}
We now estimate the \ different terms of $Z_{1}$, beginning with the terms $%
X_{1}$ and $X_{2}$ coming from $F_{0}:$%
\begin{equation*}
|X_{1}|\equiv |\int_{\Sigma _{t}}\{(\partial ^{a}N\hat{\Delta}_{g}u+\partial
^{c}N\hat{\nabla}^{a}\partial _{c}u).\hat{\nabla}_{a}u^{\prime }\}\mu _{g}|
\end{equation*}
A proof analogous to the proof of the lemma 4.2 gives 
\begin{equation}
\int_{\Sigma _{t}}|\hat{\nabla}Du|_{g}^{2}\mu _{g}\leq \int_{\Sigma _{t}}\{|%
\hat{\Delta}_{g}u|_{g}^{2}-\frac{1}{2}R(g)|Du|_{g}^{2}\}\mu _{g}.
\end{equation}
The hamiltonian constraint 2.21 implies that 
\begin{equation}
R(g)=|u^{\prime }|^{2}+|Du|_{g}^{2}+|h|_{g}^{2}-\frac{1}{2}\tau ^{2}\geq -%
\frac{1}{2}\tau ^{2}
\end{equation}
therefore 
\begin{equation}
||\hat{\nabla}Du||_{g}^{2}\leq ||\hat{\Delta}_{g}u||_{g}^{2}+\frac{1}{4}\tau
^{2}||Du||_{g}^{2}
\end{equation}
Using the estimate 6.6\ of $DN$ gives then 
\begin{equation}
|X_{1}|\leq C_{E,\lambda ,\sigma }|\tau |^{3}(\varepsilon +\varepsilon
_{1})^{4}
\end{equation}
The remaining, $X_{2},$ of the integral of $F_{0}$ is estimated as follows 
\begin{equation}
|X_{2}|=|\int_{\Sigma _{t}}(\hat{\nabla}_{a}\partial ^{c}N)\partial _{c}u)].%
\hat{\nabla}_{b}u^{\prime }\mu _{g}|\leq (||\nabla ^{2}N||_{L^{4}(g)}||\text{
}|Du|\text{ }||_{L^{4}(g)}||\hat{\nabla}u^{\prime }||_{g}
\end{equation}
The estimates of $||\nabla ^{2}N||_{L^{4}(g)}$ and $||$ $|Du|$ $%
||_{L^{4}(g)} $ given in [CB-M1], section 10.2.2, for the estimate of $Y_{2}$
applies here, due to the lemma 5.1, and give 
\begin{equation}
|X_{2}|\leq C_{E,\lambda ,\sigma }|\tau |^{3}(\varepsilon +\varepsilon
_{1})^{4}.
\end{equation}
The terms in 
\begin{equation}
X_{3}\equiv \int_{\Sigma _{t}}F_{1}.\hat{\Delta}_{g}u\mu _{g}\equiv
\int_{\Sigma _{t}}[2\partial _{c}u(h_{g}^{ac}\partial _{a}N+N\nabla
_{a}k^{ac})+2Nh_{g}^{ab}\hat{\nabla}_{a}\partial _{b}u].\hat{\Delta}_{g}u\mu
_{g}
\end{equation}
are analogous to terms found in [CB-M1] and can be estimated similarly,
giving an inequality of the form 
\begin{equation}
|X_{3}|\leq C_{E,\lambda ,\sigma }|\tau |^{3}(\varepsilon +\varepsilon
_{1})^{3}.
\end{equation}
The new terms, in our unpolarized case, are 
\begin{equation}
X_{4}\equiv \int_{\Sigma _{t}}\hat{F}_{1}.\hat{\Delta}_{g}u\mu _{g}\equiv
\int_{\Sigma _{t}}g^{ab}R_{CB,AD}\partial _{0}u^{C}\partial
_{a}u^{B}\partial _{b}u^{D}\hat{\Delta}_{g}u^{A}\mu _{g}
\end{equation}
and 
\begin{equation}
X_{5}\equiv \int_{\Sigma _{t}}\hat{F}_{0}\mu _{g}\equiv \int_{\Sigma
_{t}}R_{AB,CD}u^{\prime D}\partial _{0}u^{A}\partial ^{a}u^{B}\hat{\nabla}%
_{a}u^{\prime C}\mu _{g}
\end{equation}
We \ have here $|Riemann(G)|=4,$ therefore, using $\partial _{0}u\equiv
Nu^{\prime }$ and $0<N\leq 2,$ and the H\"{o}lder inequality: 
\begin{equation}
|X_{4}|\leq 8\int_{\Sigma _{t}}|u^{\prime }|\text{ }|Du|_{g}^{2}|\hat{\Delta}%
_{g}u|_{g}\mu _{g}\leq 8|\tau |\varepsilon _{1}||u^{\prime }||_{L^{6}(g)}||%
\text{ }|Du|^{2}\text{ }||_{L^{3}(g)}.
\end{equation}
and 
\begin{equation}
|X_{5}|\leq 8||u^{\prime }||_{L^{6}(g)}^{2}||\text{ }|Du|_{g}\text{ }%
||_{L^{6}(g)}|\tau |\varepsilon _{1}
\end{equation}
The $L^{6}(g)$ norms can be estimated as follows.\ It results from the
definitions that: 
\begin{equation}
||\text{ }|Du|_{g}\text{ }||_{L^{6}(g)}=\{\int_{\Sigma _{t}}e^{-4\lambda
}|Du|^{6}\mu _{\sigma }\}^{\frac{1}{6}}\leq e^{-\frac{2}{3}\lambda _{m}}||%
\text{ }|Du|\text{ }||_{6}
\end{equation}
while, by the Sobolev embedding theorem 
\begin{equation}
||\text{ }|Du|\text{ }||_{6}\leq C_{\sigma }(||Du||+||\text{ }D|Du|\text{ }||
\end{equation}
It holds that 
\begin{equation}
D|Du|=\frac{D|Du|^{2}}{2|Du|}=\frac{\hat{D}Du.Du}{|Du|}
\end{equation}
hence 
\begin{equation}
|D|Du|\text{ }|\leq |\hat{D}Du|
\end{equation}
and 
\begin{equation}
||\text{ }|Du|\text{ }||_{6}\leq C_{\sigma }(||Du||+||\text{ }\hat{D}Du\text{
}||)
\end{equation}
The inequalities of the lemma 5.1 and the lower bound of $\lambda _{m}$ give
then 
\begin{equation}
||\text{ }|Du|\text{ }||_{L^{6}(g)}\leq C_{\sigma }C_{\lambda }|\tau |^{%
\frac{2}{3}}(\varepsilon +\varepsilon _{1}))
\end{equation}
An analogous proof gives that 
\begin{equation}
||u^{\prime }||_{L^{6}(g)}\leq e^{\frac{1}{3}\lambda _{M}}||u^{\prime }||_{6}
\end{equation}
and 
\begin{equation}
||u^{\prime }||_{6}\leq C_{\sigma }(||u^{\prime }||+||D|u^{\prime }|\text{ }%
||),
\end{equation}
with 
\begin{equation}
|D|u^{\prime }|\text{ }|\leq |\hat{D}u^{\prime }|,\text{ \ \ }||\hat{D}%
u^{\prime }||=||\hat{D}u^{\prime }||_{g}
\end{equation}
therefore, using again the lemma 5.1 
\begin{equation}
||u^{\prime }||_{L^{6}(g)}\leq C_{\sigma }e^{\frac{1}{3}\lambda _{M}}|\tau
|(\varepsilon +\varepsilon _{1})\leq C_{\sigma }C_{\lambda }|\tau |^{\frac{2%
}{3}}(\varepsilon +\varepsilon _{1}).
\end{equation}
These estimates imply that 
\begin{equation}
|X_{4}|\leq C_{\sigma }C_{\lambda }|\tau |^{3}(\varepsilon +\varepsilon
_{1})^{3}\varepsilon _{1}
\end{equation}
and the same inequality for $X_{5}.$ We have proved the following theorem

\begin{theorem}
The second energy satisfies an equality of the form 
\begin{equation}
\frac{dE^{(1)}}{dt}-2\tau E^{(1)}(t)\leq|\tau|^{3}B_{1}.
\end{equation}
with 
\begin{equation*}
|B_{1}|\leq C_{\sigma}C_{\lambda}C_{E}(\varepsilon+\varepsilon_{1})^{3}.
\end{equation*}
\end{theorem}

\section{Corrected first energy.}

\subsection{Definition and lower bound.}

One defines as follows a corrected first energy where $\alpha$ is a
constant, which we will choose positive:

\begin{equation}
E_{\alpha}(t)=E(t)-\alpha\tau E_{c}(t),\text{ \ \ \ }E_{c}(t)\equiv
\int_{\Sigma_{t}}(u-\overset{\_}{u}).u^{\prime}\mu_{g}
\end{equation}
where we have set: 
\begin{equation}
(u-\overset{\_}{u}).u^{\prime}\equiv2(\gamma-\overset{\_}{\gamma}%
)\gamma^{\prime}+\frac{1}{2}e^{-4\gamma}(\omega-\overset{\_}{\omega}%
)\omega^{\prime}
\end{equation}
and denoted by $\overset{\_\text{ }}{f}$ the mean value on $%
(\Sigma_{t},\sigma)$ of a scalar function $f:$

\begin{equation*}
\overset{\_}{f}=\frac{1}{Vol_{\sigma}(\Sigma_{t})}\int_{\Sigma_{t}}f\mu_{%
\sigma}\equiv-\frac{1}{4\pi\chi}\int_{\Sigma_{t}}f\mu_{\sigma}
\end{equation*}
An estimate of $E_{\alpha}$ will involve second derivatives of $u,$ it
cannot alone gives a bound of the first energy $E.$

The Cauchy Schwarz inequality on ($\Sigma,g)$ and the relation between $g$
and $\sigma$ imply that:

\begin{equation}
|\int_{\Sigma_{t}}(\gamma-\overset{\_}{\gamma})\gamma^{\prime}\mu_{g}|\leq||%
\gamma-\bar{\gamma}||_{g}\parallel\gamma^{\prime}\parallel_{g}\leq
e^{\lambda_{M}}||\gamma-\bar{\gamma}||\text{ }||\gamma^{\prime}||_{g}\text{
\ }
\end{equation}
Using the Poincar\'{e} inequality and recalling that $\Lambda_{\sigma}$
denotes the first positive eigen value of the laplacian $\Delta_{\sigma}$ on
functions with mean value zero gives the majoration (recall that $\parallel
Df\parallel=\parallel Df\parallel_{g}$ if $f$ \ is a scalar function): 
\begin{equation}
|\int_{\Sigma_{t}}(\gamma-\overset{\_}{\gamma})\gamma^{\prime}\mu_{g}|\leq
e^{\lambda_{M}}\Lambda_{\sigma}^{-1/2}\parallel D\gamma\parallel
\parallel\gamma^{\prime}\parallel_{g}
\end{equation}
an analogous procedure gives, with $\gamma_{m}$ and $\gamma_{M}$ the lower
and upper bounds of $\gamma:$ 
\begin{equation}
|\int_{\Sigma_{t}}e^{-4\gamma}(\omega-\overset{\_}{\omega}%
)\omega^{\prime}\mu_{g}|\leq
e^{2(\gamma_{M}-\gamma_{m})}e^{\lambda_{M}}\Lambda_{\sigma }^{-1/2}\parallel
e^{-2\gamma}D\omega\parallel\parallel
e^{-2\gamma}\omega^{\prime}\parallel_{g}
\end{equation}
since 
\begin{equation*}
\parallel D\omega\parallel\leq e^{2\gamma_{M}}\parallel
e^{-2\gamma}D\omega\parallel.
\end{equation*}
Using the definition of the $G-$norm we see that the inequalities 9.4, 9.5
imply: 
\begin{equation}
|\int_{\Sigma_{t}}(u-\overset{\_}{u}).u^{\prime}\mu_{g}|\leq
e^{\lambda_{M}}\Lambda_{\sigma}^{-{\frac{1}{2}}}e^{2(\gamma_{M}-\gamma_{m})}%
\parallel Du\parallel\parallel u^{\prime}\parallel_{g}
\end{equation}
with (theorem 5.7) 
\begin{equation}
e^{\lambda_{M}}\leq|\tau|^{-1}\{\sqrt{2}+C_{\sigma}C_{E}C_{\lambda
}(\varepsilon+\varepsilon_{1})\}.
\end{equation}

\begin{lemma}
It holds that: 
\begin{equation*}
\gamma_{M}-\gamma_{m}\leq C_{\sigma}C_{E}C_{\lambda}\{\varepsilon
+\varepsilon_{1}\}.
\end{equation*}
\end{lemma}

\begin{proof}
We have: 
\begin{equation*}
0\leq\gamma_{M}-\gamma_{m}\leq2\parallel\gamma-\overset{\_}{\gamma}%
\parallel_{L^{\infty}}
\end{equation*}
The Sobolev imbedding theorem gives therefore that: 
\begin{equation*}
\gamma_{M}-\gamma_{m}\leq2C_{\sigma}\parallel\gamma-\overset{\_}{\gamma }%
\parallel_{H_{2}}
\end{equation*}
hence, using again the Poincar\'{e} inequality to estimate \TEXTsymbol{\vert}%
\TEXTsymbol{\vert}$\gamma-\bar{\gamma}||$, 
\begin{equation*}
\gamma_{M}-\gamma_{m}\leq2C_{\sigma}\{(\Lambda_{\sigma}^{-1}+1)\parallel
D\gamma\parallel+\parallel D^{2}\gamma\parallel\}.
\end{equation*}
It results from the definitions of the $G$ norm and of $\varepsilon$ that 
\begin{equation}
||D\gamma||^{2}\leq\frac{1}{2}||Du||^{2}\leq\varepsilon^{2}.
\end{equation}
On the other hand since $\gamma$ is a scalar function on a 2-manifold with
constant scalar curvature $-1,$\ it holds that: 
\begin{equation}
||D^{2}\gamma||^{2}=||\Delta\gamma||^{2}+\frac{1}{2}||D\gamma||^{2}
\end{equation}
We have 
\begin{equation*}
\Delta\gamma=\hat{\Delta}\gamma-\frac{1}{2}e^{-4\gamma}|D\omega|^{2}
\end{equation*}
hence 
\begin{equation}
||\Delta\gamma||\leq||\hat{\Delta}\gamma||+\frac{1}{2}||\text{ }|e^{-2\gamma
}D\omega|^{2}||
\end{equation}
with 
\begin{equation}
||\hat{\Delta}\gamma||^{2}\leq\frac{1}{2}||\hat{\Delta}u||^{2}\leq\frac{1}{2}%
e^{2\lambda_{M}}||\hat{\Delta}_{g}u||^{2}\leq C_{\lambda}\varepsilon
_{1}^{2},
\end{equation}
and, using the lemma 5.1 
\begin{equation}
||\text{ }|e^{-2\gamma}D\omega|^{2}||\leq||\text{ }|Du|^{2}||\leq C_{\sigma
}C_{\lambda}\{\varepsilon^{2}+\varepsilon\varepsilon_{1}\}.
\end{equation}
hence: 
\begin{equation}
||D^{2}\gamma||^{2}\leq
C_{E}C_{\lambda}C_{\sigma}(\varepsilon_{1}^{2}+\varepsilon^{2})
\end{equation}
Using the hypothesis H$_{E}$ we deduce from all these inequalities the
announced result.
\end{proof}

We deduce from this lemma and the elementary calculus formula: 
\begin{equation*}
e^{2(\gamma_{M}-\gamma_{m})}\leq1+2(\gamma_{M}-\gamma_{m})e^{2(\gamma
_{M}-\gamma_{m})}
\end{equation*}
that there exists an inequality of the form 
\begin{equation}
e^{2(\gamma_{M}-\gamma_{m})}\leq1+C_{E,\lambda,\sigma}(\varepsilon
+\varepsilon_{1}).
\end{equation}
\bigskip

We have proved that 
\begin{equation}
|\tau \int_{\Sigma _{t}}(u-\overset{\_}{u}).u^{\prime }\mu _{g}|\leq \sqrt{2}%
\Lambda _{\sigma }^{-{\frac{1}{2}}}\parallel Du\parallel \parallel u^{\prime
}\parallel _{g}+A_{1}
\end{equation}
with 
\begin{equation}
|A_{1}|\leq C_{E,\lambda ,\sigma }\varepsilon ^{2}(\varepsilon +\varepsilon
_{1}).
\end{equation}
Therefore: 
\begin{equation}
E_{\alpha }(t)\geq \frac{1}{2}||h||_{g}^{2}+Q_{\alpha ,\Lambda
}(x_{0},x_{1})-C_{E,\lambda ,\sigma }\varepsilon ^{2}(\varepsilon
+\varepsilon _{1}).
\end{equation}
where $\parallel Du\parallel =x_{1}$, $\parallel u^{\prime }\parallel
_{g}=x_{0},$ $x=(x_{0},x_{1})$ and $Q_{\alpha ,\Lambda }$ is the quadratic
form 
\begin{equation}
Q_{\alpha ,\Lambda }(x)\equiv \frac{1}{2}(x_{0}^{2}+x_{1}^{2})-\alpha \sqrt{2%
}\Lambda _{\sigma }^{-\frac{1}{2}}x_{0}x_{1}.
\end{equation}
The right hand side of the inequality 9.17 can be positive only if this
quadratic form is positive, that is if: 
\begin{equation}
\alpha <\frac{\Lambda _{\sigma }^{\frac{1}{2}}}{\sqrt{2}}.
\end{equation}
There exists a number $K>0$ such that: 
\begin{equation}
Q_{\alpha ,\Lambda }(x)\geq \frac{1}{2}K(x_{0}^{2}+x_{1}^{2})
\end{equation}
if the following quadratic form $Q_{K}$ is positive definite 
\begin{equation}
Q_{K}(x)\equiv (1-K)(x_{0}^{2}+x_{1}^{2})-2\alpha \sqrt{2}\Lambda _{\sigma
}^{-\frac{1}{2}}x_{0}x_{1})
\end{equation}
that is,\ under the condition 9.19\ on $\alpha ,$ 
\begin{equation}
0<K<1-\alpha \sqrt{2}\Lambda _{\sigma }^{-\frac{1}{2}}.\text{ }
\end{equation}
There will then exist a number $0<\ell <K$ such that 
\begin{equation}
E_{\alpha }(t)\geq \ell E(t)
\end{equation}
as soon as 
\begin{equation}
\varepsilon +\varepsilon _{1}\leq \eta _{3}
\end{equation}
with ($C_{\sigma ,E,\lambda }$ denotes the coefficient of this type in 9.17) 
\begin{equation}
\eta _{3}=\frac{K-\ell }{C_{\sigma ,E,\lambda }}<\frac{1-\alpha \sqrt{2}%
\Lambda _{\sigma }^{-\frac{1}{2}}}{C_{\sigma ,E,\lambda }}.
\end{equation}

\subsection{Time derivative of the corrected first energy.}

We have (recall that terms involving the shift are exact divergences which
integrate to zero, and we have set $\frac{d\tau}{dt}=\tau^{2}$): 
\begin{equation*}
\frac{dE_{\alpha}}{dt}=\frac{dE}{dt}-\alpha\tau\mathcal{R}\text{ \ \ with \
\ }\mathcal{R\equiv}\frac{dE_{c}}{dt}+\tau E_{c},
\end{equation*}
that is: 
\begin{equation}
\mathcal{R\equiv}\int_{\Sigma_{t}}\{\hat{\partial}_{0}u^{\prime}.(u-\overset{%
\_}{u})+u^{\prime}.\hat{\partial}_{0}(u-\overset{\_}{u})-N\tau u^{\prime}.(u-%
\overset{\_}{u})+\tau u^{\prime}.(u-\overset{\_}{u})\}\mu_{g}
\end{equation}
The mapping $u$ satisfies the wave map equation: 
\begin{equation}
-\hat{\partial}_{0}u^{\prime}+\hat{\nabla}^{a}(N\partial_{a}u)+\tau
Nu^{\prime}=0,
\end{equation}
therefore, performing an integration by parts where we derivate $u-\bar{u}$
as if it were a section of $E^{0,1}$ we obtain that: 
\begin{equation}
\int_{\Sigma_{t}}(\hat{\partial}_{0}u^{\prime}-N\tau u^{\prime}).(u-\overset{%
\_}{u})\mu_{g}=\int_{\Sigma_{t}}\hat{\nabla}^{a}(N\partial_{a}u).(u-\bar
{u})\mu_{g}
\end{equation}
\begin{equation}
=\int_{\Sigma_{t}}-Ng^{ab}\partial_{a}u.\hat{\partial}_{b}(u-\bar{u})\mu_{g}
\end{equation}
where 
\begin{equation}
\hat{\partial}_{b}(u^{B}-\bar{u}^{B})\equiv\partial_{b}(u^{B}-\bar{u}%
^{B})+G_{CD}^{B}\partial_{b}u^{C}(u^{D}-\bar{u}^{D})
\end{equation}
that is, due to the values of the coefficients $G_{BC}^{A}$ (remark 2.2)$:$%
\begin{equation}
\hat{\partial}_{b}(\gamma-\bar{\gamma})\equiv\partial_{b}(\gamma-\bar{\gamma 
})+\frac{1}{2}e^{-4\gamma}\partial_{b}\omega(\omega-\bar{\omega}),\text{ \ }
\end{equation}
\begin{equation}
\hat{\partial}_{b}(\omega-\bar{\omega})\equiv\partial_{b}(\omega-\bar{\omega 
})-2e^{-4\gamma}[\partial_{b}\omega(\gamma-\bar{\gamma})+\partial_{b}\gamma(%
\omega-\bar{\omega})]
\end{equation}
hence, using the expression of the metric $G:$%
\begin{equation}
\int_{\Sigma_{t}}-Ng^{ab}\partial_{a}u.\hat{\partial}_{b}(u-\bar{u})\mu
_{g}=-\int_{\Sigma_{t}}N\{|Du|_{g}^{2}-e^{-4\gamma}g^{ab}\partial_{a}\omega%
\partial_{b}\omega(\gamma-\bar{\gamma})\}\mu_{g}
\end{equation}
The non linear term can be estimated using previous results, namely (recall $%
g^{ab}\mu_{g}=\sigma^{ab}\mu_{\sigma})$: 
\begin{equation}
|\int_{\Sigma_{t}}Ne^{-4\gamma}g^{ab}\partial_{a}\omega\partial_{b}\omega(%
\gamma-\bar{\gamma})\mu_{g}|\leq2||\gamma-\bar{\gamma}||\text{ }||\,\
|e^{-2\gamma}D\omega|^{2}||.
\end{equation}
It holds that 
\begin{equation}
||\gamma-\bar{\gamma}||\leq\Lambda_{\sigma}^{-\frac{1}{2}}||D\gamma||,\text{
\ \ }||D\gamma||^{2}\leq\varepsilon^{2}
\end{equation}
and, due to the definition of the norm in $G$ and the lemma 5.1, 
\begin{equation}
||\,\ |e^{-2\gamma}D\omega|^{2}||\leq2||\,\ |Du|^{2}||\leq C_{\sigma
}C_{\lambda}(\varepsilon+\varepsilon_{1})^{2}
\end{equation}

On the other hand: 
\begin{equation}
\hat{\partial}_{0}(u-\overset{\_}{u})^{A}=\partial_{0}(u-\bar{u}%
)^{A}+G_{CD}^{A}\partial_{0}u^{C}(u-\bar{u})^{D}.
\end{equation}
A straightforward computation using the values of the coefficients $%
G_{CD}^{A}$ gives that 
\begin{equation*}
\int_{\Sigma_{t}}u^{\prime}.\hat{\partial}_{0}(u-\overset{\_}{u}%
)\mu_{g}=\int_{\Sigma_{t}}\{N|u^{\prime}|^{2}-Ne^{-4\gamma}\omega^{\prime2}(%
\gamma-\bar{\gamma})\}\mu_{g}
\end{equation*}
\begin{equation}
-\partial_{t}\bar{\gamma}\int_{\Sigma_{t}}\frac{1}{2}\gamma^{\prime}\mu
_{g}-\partial_{t}\bar{\omega}\int_{\Sigma_{t}}2e^{-4\gamma}\omega^{\prime}%
\mu_{g}.
\end{equation}

The non linear term can be estimated as before: 
\begin{equation}
\int_{\Sigma_{t}}Ne^{-4\gamma}\omega^{\prime2}(\gamma-\bar{\gamma}%
)\mu_{g}\leq2e^{2\lambda_{M}}\Lambda_{\sigma}^{-\frac{1}{2}}\varepsilon||%
\text{ }|u^{\prime}|^{2}||\leq
C_{\sigma}C_{\lambda}\varepsilon^{2}(\varepsilon +\varepsilon_{1})
\end{equation}

To bound the remaining terms we observe that for a scalar function $f$ ,
since $V_{\sigma}=-4\pi\chi$ is a constant, it holds that 
\begin{equation}
\partial_{t}\bar{f}=\frac{1}{V_{\sigma}}\partial_{t}\int_{\Sigma_{t}}f\mu_{%
\sigma}=\frac{1}{V_{\sigma}}\int_{\Sigma_{t}}\{\partial_{0}f+\nu
^{a}\partial_{a}f+\frac{1}{2}f\sigma^{ab}\partial_{t}\sigma_{ab}\}\mu_{%
\sigma}
\end{equation}
with, for the considered metric $\sigma$ 
\begin{equation}
\int_{\Sigma_{t}}\sigma^{ab}\partial_{t}\sigma_{ab}\mu_{\sigma}=0.
\end{equation}
We write $\partial_{t}\bar{f}$ under the form (recall that $f^{\prime}\equiv
N^{-1}\partial_{0}f)$ 
\begin{equation}
\partial_{t}\bar{f}=\frac{1}{V_{\sigma}}\int_{\Sigma_{t}}\{2f^{\prime
}+(N-2)f^{\prime}+\nu^{a}\partial_{a}f+\frac{1}{2}(f-\bar{f})\sigma
^{ab}\partial_{t}\sigma_{ab}\}\mu_{\sigma}
\end{equation}
with 
\begin{equation}
\frac{1}{V_{\sigma}}\int_{\Sigma_{t}}2f^{\prime}\mu_{\sigma}\equiv\frac
{\tau^{2}}{V_{\sigma}}\int_{\Sigma_{t}}f^{\prime}\mu_{g}+\frac{1}{V_{\sigma}}%
\int_{\Sigma_{t}}(2-e^{2\lambda}\tau^{2})f^{\prime}\mu_{\sigma}
\end{equation}
We deduce from these equalities that 
\begin{equation}
-\partial_{t}\bar{\gamma}\int_{\Sigma_{t}}\gamma^{\prime}\mu_{g}=-\frac
{2}{V_{\sigma}}(\int_{\Sigma_{t}}\gamma^{\prime}\mu_{g})^{2}+\frac
{1}{V_{\sigma}}X
\end{equation}
with: 
\begin{equation}
X\equiv\{\int_{\Sigma_{t}}[(2-e^{2\lambda}\tau^{2}+N-2)\gamma^{\prime}+\nu
^{a}\partial_{a}\gamma+\frac{1}{2}(\gamma-\bar{\gamma})\sigma^{ab}\partial
_{t}\sigma_{ab}]\mu_{\sigma}\}\{\int_{\Sigma_{t}}\gamma^{\prime}\mu_{g}\}
\end{equation}
\ The equality 9.44 implies the inequality: 
\begin{equation}
-\partial_{t}\bar{\gamma}\int_{\Sigma_{t}}\gamma^{\prime}\mu_{g}\leq\frac
{1}{V_{\sigma}}X.
\end{equation}
All the terms in $X$ are non linear in the energies and can be estimated.
Indeed: 
\begin{equation}
|\int_{\Sigma_{t}}\gamma^{\prime}\mu_{g}|\leq V_{g}^{\frac{1}{2}%
}||\gamma^{\prime}||_{g}\leq e^{\lambda_{M}}V_{\sigma}||\gamma^{\prime}||_{g}
\end{equation}
while (theorems 5.7 and corollary 6.4) 
\begin{equation}
|2-e^{2\lambda}\tau^{2}+N-2|\leq C_{E}C_{\lambda}C_{\sigma}(\varepsilon
+\varepsilon_{1})^{2}
\end{equation}
and 
\begin{equation}
\int_{\Sigma_{t}}|\gamma^{\prime}|\mu_{\sigma}\leq V_{\sigma}^{\frac{1}{2}%
}||\gamma^{\prime}||\leq V_{\sigma}^{\frac{1}{2}}e^{-\lambda_{m}}||\gamma^{%
\prime}||_{g},
\end{equation}
Also 
\begin{equation}
||\nu||\leq C_{E}C_{\sigma}|\tau|(\varepsilon+\varepsilon_{1})
\end{equation}
and 
\begin{equation}
||D\gamma||=||D\gamma||_{g}\leq\varepsilon
\end{equation}

Using section 7.1 we find that: 
\begin{equation}
|\sigma^{ab}\partial_{t}\sigma_{ab}|\leq C_{\sigma}C_{E}|\tau|(\varepsilon
+\varepsilon_{1}).
\end{equation}

The same type of inequalities applies to the scalar function $\partial_{t}%
\bar{\omega},$ but we must now use also the identities 
\begin{equation*}
e^{-2\bar{\gamma}}\int_{\Sigma_{t}}\omega^{\prime}\mu_{g}\equiv\int
_{\Sigma_{t}}e^{-2\gamma}\omega^{\prime}e^{2(\gamma-\bar{\gamma})}\mu
_{g}\equiv\int_{\Sigma_{t}}\{e^{-2\gamma}\omega^{\prime}+(e^{2(\gamma -\bar{%
\gamma})}-1)e^{-2\gamma}\omega^{\prime}\}\mu_{g}
\end{equation*}
\begin{equation*}
e^{2\bar{\gamma}}\int_{\Sigma_{t}}e^{-4\gamma}\omega^{\prime}\mu_{g}\equiv
\int_{\Sigma_{t}}e^{-2\gamma}\omega^{\prime}e^{2(\bar{\gamma}-\gamma)}\mu
_{g}\equiv\int_{\Sigma_{t}}\{e^{-2\gamma}\omega^{\prime}+(e^{-2(\gamma -\bar{%
\gamma})}-1)e^{-2\gamma}\omega^{\prime}\}\mu_{g}
\end{equation*}
to obtain an inequality which bounds $-\partial_{t}\bar{\omega}\int
_{\Sigma_{t}}e^{-4\gamma}\omega^{\prime}\mu_{g}$ with higher order terms in
the energies, using the bound 
\begin{equation}
|e^{\pm2(\gamma-\bar{\gamma})}-1|\leq2|\gamma-\bar{\gamma}|e^{2|\gamma -\bar{%
\gamma}|}\leq C_{E,\lambda,\sigma}(\varepsilon+\varepsilon_{1}).
\end{equation}
We have proved that 
\begin{equation}
\mathcal{R}\leq\int_{\Sigma_{t}}\{-N|Du|^{2}+N|u^{\prime}|^{2}+\tau
u^{\prime }.(u-\bar{u})\}\mu_{g}+A_{2}
\end{equation}
with: 
\begin{equation}
|A_{2}|\leq C_{E,\lambda,\sigma}(\varepsilon+\varepsilon_{1})^{3}.
\end{equation}

\begin{theorem}
There exist numbers $\alpha>0$ and $k>0$ such that 
\begin{equation}
\frac{dE_{\alpha}}{dt}-k\tau E_{\alpha}\leq|\tau A|
\end{equation}
with 
\begin{equation}
|\tau A|\leq|\tau|C_{E,\lambda,\sigma}(\varepsilon+\varepsilon_{1})^{3}.
\end{equation}

1. If $\Lambda _{\sigma }>\frac{1}{8}$ the best choice is 
\begin{equation}
\alpha =\frac{1}{4},\ k=1
\end{equation}

\ 2. $\Lambda _{\sigma }\leq \frac{1}{8}.$ Then $\alpha $ and $k$ are such
that: 
\begin{equation}
0<\alpha <\frac{4}{8+\Lambda _{\sigma }^{-1}}\leq \frac{1}{4},\text{ \ \ }%
0<k<1-\frac{1-4\alpha }{(1-2\Lambda _{\sigma }^{-1}\alpha ^{2})^{\frac{1}{2}}%
}
\end{equation}

A number $\alpha$ satisfying the conditions of the above theorem is also
such that $0<\alpha<\frac{\Lambda_{\sigma}^{\frac{1}{2}}}{\sqrt{2}}.$
\end{theorem}

\begin{proof}
using 10.54 and the expression 4.8\ of $\frac{dE}{dt}$ we find that: 
\begin{equation}
\frac{dE_{\alpha}}{dt}\leq\tau\int_{\Sigma_{t}}\{|h|^{2}+(1-2\alpha
)|u^{\prime}|^{2}+2\alpha|Du|_{g}^{2}]-\alpha\tau u^{\prime}.(u-\bar{u}%
)\}\mu_{g}+|\tau A|
\end{equation}
where 
\begin{equation}
A\equiv\alpha A_{1}+A_{2}.
\end{equation}
with 
\begin{equation}
A_{2}\equiv\int_{\Sigma_{t}}\frac{1}{2}(N-2)[|h|^{2}+(1-2\alpha)|u^{\prime
}|^{2}+2\alpha|Du|_{g}^{2}]\mu_{g}
\end{equation}
We deduce from the corollary 6.4 ($L^{\infty}$ estimate of $N-2)$ that $%
A_{2} $ satisfies the same type of estimate than $A_{1},$ hence: 
\begin{equation}
|\tau
A|\leq|\tau|C_{E}C_{\lambda}C_{\sigma}(\varepsilon+\varepsilon_{1})^{3}.
\end{equation}
We look for a positive number $k$ such that the difference $\frac{dE_{\alpha}%
}{dt}-k\tau E_{\alpha}$ can be estimated with higher order terms in the
energies. We deduce from 9.60 that: 
\begin{equation}
\frac{dE_{\alpha}}{dt}-k\tau E_{\alpha}\leq\tau\{||h||_{g}{}^{2}+(1-2\alpha-%
\frac{k}{2})||u^{\prime}||_{g}{}^{2}+(2\alpha-\frac{k}{2})||Du||_{g}^{2}
\end{equation}
\begin{equation}
+\alpha\int_{\Sigma_{t}}|\tau|(1-k)u^{\prime}.(u-\bar{u})\mu_{g}\}+|\tau A|
\end{equation}
We have seen that: 
\begin{equation}
|\tau\int_{\Sigma_{t}}u^{\prime}.(u-\overset{\_}{u})\mu_{g}|\leq\sqrt
{2}\Lambda_{\sigma}^{-1/2}||u^{\prime}||_{g}||Du||_{g}+A_{1},
\end{equation}

Since $\tau<0$, it will hold that 
\begin{equation}
\frac{dE_{\alpha}}{dt}-k\tau E_{\alpha}\leq|\tau A|,\text{ \ \ }A\equiv
A_{1}+A_{2}+A_{3}.
\end{equation}
if the quadratic form 
\begin{equation}
Q_{\alpha,k}(x)\equiv(1-2\alpha-\frac{k}{2})x_{0}^{2}+(2\alpha-\frac{k}{2}%
)x_{1}^{2}-\alpha(1-k)\sqrt{2}\Lambda_{\sigma}^{-1/2}x_{0}x_{1}
\end{equation}
is non negative.

The quadratic form $Q_{\alpha,k}$ is non negative if: 
\begin{equation}
k\leq4\alpha,\text{ \ and \ \ \ }k\leq2(1-2\alpha)
\end{equation}
and $k$ is such that its discriminant is negative, that is: 
\begin{equation}
2\alpha^{2}\Lambda_{\sigma}^{-1}(1-k)^{2}-4(2\alpha-\frac{k}{2})(1-2\alpha -%
\frac{k}{2})\leq0
\end{equation}
The inequalities 9.69 imply 
\begin{equation}
k\leq1,
\end{equation}
The inequality 9.70 reads 
\begin{equation}
(1-2\Lambda_{\sigma}^{-1}\alpha^{2})k^{2}-(1-2\Lambda_{\sigma}^{-1}\alpha
^{2})2k-2\Lambda_{\sigma}^{-1}\alpha^{2}+8\alpha(1-2\alpha)>0
\end{equation}
We have already supposed that $1-2\Lambda_{\sigma}^{-1}\alpha^{2}>0,$ the
inequality above is therefore equivalent to: 
\begin{equation}
k^{2}-2k+1-\frac{(1-4\alpha)^{2}}{(1-2\Lambda_{\sigma}^{-1}\alpha^{2})}>0
\end{equation}
that is 
\begin{equation}
k<1-\frac{1-4\alpha}{(1-2\Lambda_{\sigma}^{-1}\alpha^{2})^{\frac{1}{2}}}
\end{equation}
There will exist such a $k>0$ if 
\begin{equation}
\frac{1-4\alpha}{(1-2\Lambda_{\sigma}^{-1}\alpha^{2})^{\frac{1}{2}}}<1
\end{equation}
Since $\alpha>0$ this inequality reduces to: 
\begin{equation}
-2\Lambda_{\sigma}^{-1}\alpha-16\alpha+8>0.
\end{equation}
i.e. 
\begin{equation}
\alpha<\frac{4}{8+\Lambda_{\sigma}^{-1}}
\end{equation}
We remark that this inequality imposes the hypothesis first made on $\alpha, 
$ since elementary calculus shows that, for any $\Lambda,$ it holds that: 
\begin{equation}
\frac{\Lambda^{\frac{1}{2}}}{\sqrt{2}}\leq\frac{4}{8+\Lambda^{-1}},
\end{equation}
the inequality being attained only for $\Lambda=\frac{1}{8}.$

We distinguish two cases

1. $\Lambda_{\sigma}>\frac{1}{8}.$ In this case it is possible to take $%
\alpha=\frac{1}{4},\ k=1$ and obtain immediately 
\begin{equation}
\frac{dE_{\frac{1}{4}}}{dt}-\tau E_{\frac{1}{4}}\leq|\tau A|.
\end{equation}

2. $\Lambda_{\sigma}\leq\frac{1}{8}.$ We have then: 
\begin{equation}
\frac{4}{8+\Lambda_{\sigma}^{-1}}\leq\frac{1}{4}.
\end{equation}
We choose $\alpha$ such that it satisfies the inequality 9.77, which implies
in this case $\alpha<\frac{1}{4},$ and then $k>0$ such that it satisfies
9.74.
\end{proof}

\section{Corrected second energy.}

We define a \textbf{corrected second energy} $E_{\alpha}^{(1)}$ by the
formula 
\begin{equation}
E_{\alpha}^{(1)}(t)=E^{(1)}(t)+\alpha\tau E_{c}^{(1)}(t),\text{ \ \ }%
E_{c}^{(1)}(t)\equiv\int_{\Sigma_{t}}\hat{\Delta}_{g}u.u^{\prime}\mu_{g}
\end{equation}

\subsection{Lower bound.}

We have, according to previous notations, 
\begin{equation}
\hat{\Delta}_{g}u.u^{\prime}\equiv2\Delta_{g}\gamma\gamma^{\prime}+\frac{1}{4%
}e^{-4\gamma}\Delta_{g}\omega\omega^{\prime}+b_{1}
\end{equation}
with 
\begin{equation}
b_{1}\equiv e^{-4\gamma}g^{ab}(\partial_{a}\omega\partial_{b}\omega
\gamma^{\prime}-\partial_{a}\gamma\partial_{b}\omega\omega^{\prime}).
\end{equation}
hence, using the lemma 5.2: 
\begin{equation}
B_{1}\equiv|\int_{\Sigma_{t}}b_{1}\mu_{g}|\leq|\tau|C_{E,\lambda,\sigma
}(\varepsilon+\varepsilon_{1})^{3}.
\end{equation}
The Cauchy Schwarz inequality and the Poincar\'{e} inequality ($\bar{\gamma }%
^{\prime}$ is a constant on $\Sigma_{t}$ and on a compact manifold $%
\int_{\Sigma_{t}}\Delta_{g}\gamma\mu_{g}=0)$ give that: 
\begin{equation}
|\int_{\Sigma_{t}}\Delta_{g}\gamma\gamma^{\prime}\mu_{g}|=|\int_{\Sigma_{t}}%
\Delta_{g}\gamma(\gamma^{\prime}-\bar{\gamma}^{\prime})\mu_{g}|\leq
e^{\lambda_{M}}\Lambda_{\sigma}^{-1/2}||\Delta_{g}\gamma||_{g}||D\gamma
^{\prime}||
\end{equation}
while 
\begin{align*}
|\int_{\Sigma_{t}}(\Delta_{g}\omega)e^{-4\gamma}\omega^{\prime}\mu_{g}| &
=|\int_{\Sigma_{t}}\Delta_{g}\omega(e^{-4\gamma}\omega^{\prime}-\overline
{e^{-4\gamma}\omega^{\prime}})\mu_{g}| \\
& \leq e^{\lambda_{M}}\Lambda_{\sigma}^{-1/2}||\Delta_{g}\omega
||_{g}||D(e^{-4\gamma}\omega^{\prime})||
\end{align*}
It holds that 
\begin{align}
||D(e^{-4\gamma}\omega^{\prime})|| & =||e^{-4\gamma}(D\omega^{\prime
}-4D\gamma\omega^{\prime})|| \\
& \leq e^{-2\gamma_{m}}(||e^{-2\gamma}D\omega^{\prime})||+4||e^{-2\gamma
}\omega^{\prime}||_{4}||D\gamma||_{4}
\end{align}
while 
\begin{equation}
||\Delta_{g}\omega||_{g}\leq
e^{2\gamma_{M}}||e^{-2\gamma}\Delta_{g}\omega||_{g}
\end{equation}
Using the bound (lemma 9.1)\ of $\gamma_{M}-\gamma_{m}$ and the inequalities
on the $L^{4}$ norms \TEXTsymbol{\vert}\TEXTsymbol{\vert}.\TEXTsymbol{\vert}%
\TEXTsymbol{\vert}$_{4}$ (lemma 5.1), we find an inequality of the form: 
\begin{equation*}
|\int_{\Sigma_{t}}(\Delta_{g}\omega)e^{-4\gamma}\omega^{\prime}\mu_{g}||\leq
e^{\lambda_{M}}\Lambda_{\sigma}^{-1/2}||e^{-2\gamma}\Delta_{g}\omega
||_{g}||e^{-2\gamma}D\omega^{\prime})||+B_{2}
\end{equation*}
where $B_{2}$ satisfies an inequality of the same type as $B_{1}.$ We have
shown that 
\begin{equation*}
|\int_{\Sigma_{t}}\Delta_{g}u.u^{\prime}\mu_{g}|\leq
e^{\lambda_{M}}\Lambda_{\sigma}^{-1/2}||\Delta_{g}u||_{g}||Du^{%
\prime}||+B_{1}+B_{2}.
\end{equation*}

The estimates of the lemma 5.1 and the inequalities 
\begin{equation}
||\Delta_{g}u||_{g}\leq||\hat{\Delta}_{g}u||_{g}+||\text{ }|Du|^{2}||_{g},%
\text{ \ \ }||Du^{\prime}||\leq||\hat{\nabla}u^{\prime}||+||\text{ }%
|Du|^{2}||_{g}^{\frac{1}{2}}||\text{ }|u^{\prime}|^{2}||_{g}^{\frac{1}{2}}
\end{equation}
give 
\begin{equation*}
e^{\lambda_{M}}\Lambda_{\sigma}^{-1/2}||\Delta_{g}u||_{g}||Du^{\prime}||\leq
e^{\lambda_{M}}\Lambda_{\sigma}^{-1/2}||\hat{\Delta}_{g}u||_{g}||\hat{\nabla 
}u^{\prime}||+B_{3}
\end{equation*}
with 
\begin{equation}
B_{3}\leq|\tau|C_{E,\lambda,\sigma}(\varepsilon+\varepsilon_{1})^{3}.
\end{equation}
Using the estimate 6.23 of $e^{\lambda_{M}}|\tau|-\sqrt{2}$ we have: 
\begin{equation}
|\int_{\Sigma_{t}}\Delta_{g}u.u^{\prime}\mu_{g}|\leq\sqrt{2}%
|\tau|^{-1}\Lambda_{\sigma}^{-1/2}||\hat{\Delta}_{g}u||_{g}||\hat{\nabla}%
u^{\prime }||+B_{1}+B_{2}+B_{3}+B_{4}.
\end{equation}
with 
\begin{equation*}
B_{4}=|e^{\lambda_{M}}-\sqrt{2}|\tau|^{-1}|\Lambda_{\sigma}^{-1/2}||\hat{%
\Delta}_{g}u||_{g}||\hat{\nabla}u^{\prime}||\leq|\tau|C_{\sigma
}(\varepsilon+\varepsilon_{1})^{3}
\end{equation*}
We deduce from these estimates, with $Q_{\alpha,\Lambda}$ the same quadratic
form as in 9.18 but with $y_{1}\equiv|\tau|^{-1}||\hat{\Delta}_{g}u||_{g}$, $%
y_{0}\equiv|\tau|^{-1}||\hat{\nabla}u^{\prime}||,$ that: 
\begin{equation}
\tau^{-2}E_{\alpha}^{(1)}(t)\geq Q_{\alpha,\Lambda}(y_{0},y_{1})-C_{E,\sigma
,\lambda}(\varepsilon+\varepsilon_{1})^{3}
\end{equation}

\begin{theorem}
If $\alpha$ is chosen satisfying 9.19 there exists a number $\eta_{4}>0$ and 
$L>0$ such that 
\begin{equation}
E_{\alpha}+\tau^{-2}E_{\alpha}^{(1)}\geq
L(\varepsilon^{2}+\varepsilon_{1}^{2})
\end{equation}
as soon as 
\begin{equation}
\varepsilon+\varepsilon_{1}\leq\eta_{4}.
\end{equation}
\end{theorem}

\begin{proof}
We have found that 
\begin{equation}
\psi (t)\equiv E_{\alpha }(t)+\tau ^{-2}E_{\alpha }^{(1)}(t)\geq Q_{\alpha
,\Lambda }(y,x)-(A+B)
\end{equation}
where $Q_{\alpha ,\Lambda }(x,y)$ is the quadratic form 
\begin{equation}
Q_{\alpha ,\Lambda }(x,y)\equiv Q_{\alpha ,\Lambda }(x)+Q_{\alpha ,\Lambda
}(y).
\end{equation}
and $A+B$ admits a bound of the form 
\begin{equation}
|A+B|\leq C_{E,\sigma ,\lambda }(\varepsilon ^{2}+\varepsilon _{1}^{2})^{%
\frac{3}{2}}.
\end{equation}
We have 
\begin{equation}
Q_{\alpha ,\Lambda }(x,y)>K(\varepsilon ^{2}+\varepsilon _{1}^{2})\equiv 
\frac{1}{2}K(x_{0}^{2}+x_{1}^{2}+y_{0}^{2}+y_{1}^{2})
\end{equation}
if the quadratic form $Q_{K}$ defined in the section 9.1 is positive
definite. The conditions on $\alpha $ and the corresponding limitation on $K$
are the same as in the section 9.1, and the proof continues along the same
line.
\end{proof}

\subsection{Decay of the second corrected energy.}

We have (recall $\frac{d\tau}{dt}=\tau^{2}):$%
\begin{equation*}
\frac{dE_{\alpha}^{(1)}}{dt}\equiv\frac{dE^{(1)}}{dt}+\alpha\tau \mathcal{R}%
^{(1)}\text{ \ \ }\mathcal{R}^{(1)}\equiv\frac{d}{dt}E_{c}^{(1)}+\tau
E_{c}^{(1)}
\end{equation*}
that is: 
\begin{equation}
\mathcal{R}^{(1)}=\int_{\Sigma_{t}}\{\hat{\partial}_{0}\hat{\Delta}%
_{g}u.u^{\prime}+\hat{\Delta}_{g}u.(\hat{\partial}_{0}u^{\prime}-N\tau.u^{%
\prime}+\tau u^{\prime})\}\mu_{g}.
\end{equation}
We have found in lemma 4.1 that 
\begin{equation}
\hat{\partial}_{0}\hat{\Delta}_{g}u^{A}\equiv g^{ab}\hat{\nabla}_{a}\hat{%
\partial}_{0}\partial_{b}u^{A}+N\tau\hat{\Delta}_{g}u^{A}+F_{1}^{A}+\hat{F}%
_{1}^{A},
\end{equation}
with 
\begin{equation*}
\hat{F}_{1}^{A}\equiv g^{ab}R_{CB}{}^{A}{}_{D}\partial_{0}u^{C}\partial
_{a}u^{B}\partial_{b}u^{D}
\end{equation*}
and 
\begin{equation}
F_{1}^{A}\equiv2\partial_{c}u^{A}(h_{g}^{ac}\partial_{a}N+N%
\nabla_{a}k^{ac})+2Nh_{g}^{ab}\hat{\nabla}_{a}\partial_{b}u^{A}
\end{equation}
that is, using the equation 
\begin{equation*}
^{(3)}R_{0}^{c}\equiv-N\nabla_{a}k^{ac}=\partial_{0}u.\partial^{c}u
\end{equation*}
\begin{equation}
F_{1}^{A}\equiv2\partial_{c}u^{A}(h_{g}^{ac}\partial_{a}N-\partial
_{0}u.\partial^{c}u)+2Nh_{g}^{ab}\hat{\nabla}_{a}\partial_{b}u^{A}.
\end{equation}
Partial integration gives, using also the identity $\hat{\partial}%
_{0}\partial_{b}u\equiv\hat{\nabla}_{b}\partial_{0}u\equiv\hat{\nabla}%
_{b}(Nu^{\prime}),$ 
\begin{equation}
\int_{\Sigma_{t}}(\hat{\partial}_{0}\hat{\Delta}_{g}u).u^{\prime}\mu_{g}=%
\int_{\Sigma_{t}}\{-N|\hat{\nabla}u^{\prime}|_{g}^{2}-\partial^{a}N\hat{%
\nabla}_{a}u^{\prime}.u^{\prime}+N\tau\hat{\Delta}_{g}u.u^{\prime }+(F_{1}+%
\hat{F}_{1}).u^{\prime}\}\mu_{g}
\end{equation}

On the other hand, if $u$ satisfies the equation 
\begin{equation}
-\hat{\partial}_{0}u^{\prime}+\hat{\nabla}^{a}(N\partial_{a}u)+\tau
Nu^{\prime}=0
\end{equation}
it holds that: 
\begin{equation*}
\int_{\Sigma_{t}}\hat{\Delta}_{g}u.(\hat{\partial}_{o}u^{\prime}-\tau
Nu^{\prime})\mu_{g}=\int_{\Sigma_{t}}\{N|\hat{\Delta}_{g}u|^{2}+\partial
^{a}N\partial_{a}u.\hat{\Delta}_{g}u\}\mu_{g}
\end{equation*}
We have found that: 
\begin{equation}
\mathcal{R}^{(1)}=\int_{\Sigma_{t}}\{-N|\hat{D}u^{\prime}|^{2}+N|\hat{\Delta 
}_{g}u|^{2}+\tau(N+1)\hat{\Delta}_{g}u.u^{\prime}\}\mu_{g}+\mathcal{\tilde{R}%
}^{(1)}
\end{equation}
with 
\begin{equation}
\mathcal{\tilde{R}}^{(1)}\equiv\int_{\Sigma_{t}}\{-\partial^{a}N\hat{\nabla }%
_{a}u^{\prime}.u^{\prime}+(F_{1}+\hat{F}_{1}).u^{\prime}+\partial
^{a}N\partial_{a}u.\hat{\Delta}_{g}u\}\mu_{g}.
\end{equation}
Using the expression of $\frac{dE^{(1)}}{dt}$ we find that: 
\begin{equation*}
\frac{dE_{\alpha}^{(1)}}{dt}=\tau\int_{\Sigma_{t}}\{N(2-2\alpha)J_{0}+N(2%
\alpha+1)J_{1}+(N+1)\alpha\tau\hat{\Delta}_{g}u.u^{\prime}]\}\mu_{g}+Z_{1}+%
\alpha\tau\mathcal{\tilde{R}}^{(1)}
\end{equation*}
which implies:

\begin{equation*}
\frac{dE_{\alpha}^{(1)}}{dt}-(2+k)\tau E_{\alpha}^{(1)}=\tau\{\int_{\Sigma
_{t}}\{(2N-2-k-2\alpha N)J_{0}+(2\alpha N+N-2-k)J_{1}
\end{equation*}
\begin{equation}
+\alpha\tau(N+1-2-k)\hat{\Delta}_{g}u.u^{\prime}\}\mu_{g}+Z_{1}+\alpha \tau%
\mathcal{\tilde{R}}^{(1)}
\end{equation}
which we write:

\begin{equation*}
\frac{dE_{\alpha}^{(1)}}{dt}-(2+k)\tau E_{\alpha}^{(1)}\leq\tau\{\int
_{\Sigma_{t}}\{(2-k-4\alpha)J_{0}+(4\alpha-k)J_{1}
\end{equation*}
\begin{equation}
+\alpha\tau(1-k)\hat{\Delta}_{g}u.u^{\prime}\}\mu_{g}+Z_{1}+\alpha \tau%
\mathcal{\tilde{R}}^{(1)}+Z_{2}.
\end{equation}
with 
\begin{equation}
Z_{2}\equiv\tau\int_{\Sigma_{t}}\{(N-2)(2-2\alpha)J_{0}+(N-2)(2\alpha
+1)J_{1}+(N-2)\alpha\tau\hat{\Delta}_{g}u.u^{\prime}]\}\mu_{g}.
\end{equation}
We have found in section 8.2 that 
\begin{equation}
|Z_{1}|\leq|\tau|^{3}C_{E,\lambda,\sigma}(\varepsilon+\varepsilon_{1})^{3}
\end{equation}
It results immediately from the estimate of $N-2,$ section 6.2, that $Z_{2}$
satisfies an inequality of the same type.

Some terms of $\mathcal{\tilde{R}}^{(1)}$ are bounded using the $L^{\infty
}(g)$ estimate 6.6\ of $DN,$ which gives that 
\begin{equation}
|\int_{\Sigma_{t}}\{-\partial^{a}N\hat{\nabla}_{a}u^{\prime}.u^{\prime
}+\partial^{a}N\partial_{a}u.\hat{\Delta}_{g}u\}\mu_{g}|\leq\tau
^{2}C_{E,\sigma,\lambda}\varepsilon^{2}\varepsilon_{1}(\varepsilon
+\varepsilon_{1}).
\end{equation}
The estimate of the remaining ones uses similar techniques as those of
section 9 and lead to an inequality of the form 
\begin{equation}
|\mathcal{\tilde{R}}^{(1)}|\leq\tau^{2}C_{E,\sigma,\lambda}(\varepsilon
+\varepsilon_{1})^{4}.
\end{equation}

\begin{theorem}
Under the conditions on $\alpha$ and $k$ given in the theorem 9.2 the
following inequality holds: 
\begin{equation}
\frac{dE_{\alpha}^{(1)}}{dt}-(2+k)\tau E_{\alpha}^{(1)}\leq|\tau|^{3}B
\end{equation}
with 
\begin{equation*}
|B|\leq C_{E,\lambda,\sigma}(\varepsilon+\varepsilon_{1})^{3}.
\end{equation*}
\end{theorem}

\begin{proof}
We have seen that (10.11) 
\begin{equation}
|\int_{\Sigma_{t}}\tau\hat{\Delta}_{g}u.u^{\prime}\mu_{g}|\leq\sqrt{2}%
\Lambda_{\sigma}^{-\frac{1}{2}}||\hat{\Delta}_{g}u||_{g}\text{ }||\hat{%
\nabla }u^{\prime}||_{g}+Z_{3}
\end{equation}
with 
\begin{equation}
|Z_{3}|\leq\tau^{2}C_{E,\sigma,\lambda}(\varepsilon+\varepsilon_{1})^{3}.
\end{equation}
Therefore we deduce from 10.28 and the definition of $y_{0},y_{1}$ that 
\begin{equation*}
\frac{dE_{\alpha}^{(1)}}{dt}-(2+k)\tau
E_{\alpha}^{(1)}\leq\tau^{3}Q_{\alpha,k}^{(1)}(y)+|\tau|^{3}B
\end{equation*}
with 
\begin{equation}
B\equiv Z_{1}+\alpha\tau\mathcal{\tilde{R}}^{(1)}+Z_{2}+|\tau|\alpha Z_{3},%
\text{ \ \ }|B|\leq C_{E,\sigma,\lambda}(\varepsilon^{2}+\varepsilon
_{1}^{2})^{\frac{3}{2}}
\end{equation}
and 
\begin{equation}
Q_{\alpha,k}^{(1)}(y)\equiv(1-\frac{k}{2}-2\alpha)y_{0}^{2}+(2\alpha-\frac
{k}{2})y_{1}^{2}+\sqrt{2}\Lambda_{\sigma}^{-\frac{1}{2}}\alpha\tau
(1-k)y_{0}y_{1}.
\end{equation}

This quadratic form in $y$ is non negative under the same conditions as the
form $Q_{\alpha,k}(x).$ The conclusion follows, since $\tau<0$.
\end{proof}

\section{Decay of the total energy.}

We make the following a priori hypothesis, for all $t\geq t_{0}$ for which
the considered quantities exist

\begin{itemize}
\item  \textbf{Hypothesis H}$_{\sigma}$ : 1. The $t$ dependent numbers $%
C_{\sigma}$ are uniformly bounded by a constant $M$.
\end{itemize}

2. There exist $\Lambda>0$ such that $\Lambda_{\sigma}\geq\Lambda.$

\begin{itemize}
\item  We choose $\alpha$ such that 
\begin{equation}
\alpha=\frac{1}{4}\text{ \ \ if \ \ }\Lambda>\frac{1}{8},\text{ \ }\alpha<%
\frac{4}{8+\Lambda^{-1}}\leq\frac{1}{4}\ \ \text{if}\ \ \Lambda \leq\frac{1}{%
8}.
\end{equation}

\item  \textbf{Hypotheses } \textbf{H}$_{E}^{\eta}:$ The $t$ dependent
energies $\varepsilon^{2}$ and $\varepsilon_{1}^{2}$ having been supposed
bounded by a number $c_{E}$ we suppose, moreover that they satisfy the
inequalities 5.25, 6.3, 9.25, 10.14.

We have seen (theorem 5.7) that \ the hypothesis H$_{\lambda}$ is then
satisfied.
\end{itemize}

We denote by $M_{i}$ any given positive number dependent on the bounds of
these H's hypothesis but independent of $t$.

We have defined $\psi(t)$ to be the \textbf{total corrected energy} namely: 
\begin{equation*}
\psi(t)\equiv E_{\alpha}(t)+\tau^{-2}E_{\alpha}^{(1)}
\end{equation*}
We have seen (10.13) that $\psi(t)$\ bounds the total energy $\phi
(t)\equiv\varepsilon^{2}+\varepsilon_{1}^{2}$ by an inequality of the form: 
\begin{equation}
\phi(t)\equiv\varepsilon^{2}+\varepsilon_{1}^{2}\leq M_{0}\psi(t),\text{ \ \ 
}M_{0}=L^{-1}.
\end{equation}
\ 

\begin{lemma}
Under the hypotheses H$_{\sigma}$ and H$_{E}^{\eta}$ the function $\psi$
satisfies a differential inequality of the form 
\begin{equation}
\frac{d\psi}{dt}\leq-\frac{k}{t}(\psi-M_{1}\psi^{3/2})
\end{equation}
\end{lemma}

\begin{proof}
The inequalities 9.56 and 10.33 together with the choice $\tau=-\frac{1}{t}, 
$ and the bound 10.13.
\end{proof}

\begin{theorem}
Under the hypotheses H$_{\sigma}$ and H$_{E}^{\eta}$ there exists a number $%
k>0$ such that the total energy $E_{tot}(t)\equiv\phi(t)\equiv\varepsilon
^{2}+\varepsilon_{1}^{2}$ satisfies an estimate of the form 
\begin{equation}
t^{k}\phi(t)\leq M_{2}\phi(t_{0})
\end{equation}
if it is small enough initially.
\end{theorem}

\begin{proof}
We suppose that $\psi _{0}\equiv \psi (t_{0})$ satisfies 
\begin{equation}
\psi _{0}^{1/2}<\frac{1}{M_{1}},\text{ \ \ for instance \ \ }\psi
_{0}^{1/2}\leq \frac{1}{M_{1}}
\end{equation}
Then $\psi $ starts decreasing, continues to decrease as long as it exists,
therefore $(\psi -M_{1}\psi ^{3/2})>0$ and the inequality 11.3 is equivalent
to 
\begin{equation*}
\frac{dz}{z-M_{1}z^{2}}+\frac{k}{2}\frac{dt}{t}\leq 0,\text{ \ \ with \ \ }%
\psi =z^{2}.
\end{equation*}
This inequality gives by integration: 
\begin{equation*}
\log \{\frac{z}{(1-M_{1}z)z_{0}}\frac{(1-M_{1}z_{0})}{z_{0}}\}+\frac{1}{2}%
k\log \frac{t}{t_{0}}\leq 0\text{ }
\end{equation*}
equivalently 
\begin{equation}
\{\frac{z}{(1-M_{1}z)}\frac{(1-M_{1}z_{0})}{z_{0}}\}\{\frac{t}{t_{0}}\}^{%
\frac{1}{2}k}\leq 1
\end{equation}
and, a fortiori, 
\begin{equation}
t^{k}\psi \leq \frac{t_{0}^{k}\psi _{0}}{(1-M_{1}z_{0})^{2}}\leq
4t_{0}^{k}\psi _{0}.
\end{equation}
Hence, using 11.2, the \textbf{decay estimate,} with $M_{2}=4t_{0}^{k}M_{0},$%
\begin{equation}
t^{k}\phi (t)\leq M_{2}\phi _{0}.
\end{equation}
\end{proof}

\section{Teichm\"{u}ller parameters.}

Instead of considering as in [CB-M1], [CB-M2] the Dirichlet energy of the
metric $\sigma$ we use directly the estimate 7.7 of $dQ/dt$ which we now
write, using 11.8: 
\begin{equation}
|\frac{dQ}{dt}|\leq C_{\sigma,E}t^{-(1+\frac{k}{2})}\phi(t_{0})^{1/2}
\end{equation}
Therefore:

\begin{theorem}
There exists $M_{3}$ such that 
\begin{equation}
|Q(t)-Q(t_{0})|\leq M_{3}\phi(t_{0})^{1/2}.
\end{equation}
\end{theorem}

\section{Global existence.}

\begin{theorem}
\ Let ($\sigma_{0},q_{0})\in C^{\infty}(\Sigma_{0})$ and ($u_{0},\overset{.}{%
u}_{0})\in H_{2}(\Sigma_{0},\sigma_{0})\times H_{1}(\Sigma_{0},\sigma _{0})$
be initial data for the Einstein equations with U(1) isometry group on the
initial manifold $M_{0}\equiv\Sigma_{0}\times U(1)$ , with $\Sigma_{0}$
compact, orientable and of genus greater than one, $\sigma_{0}$ chosen such
that $R(\sigma_{0})=-1.$ Suppose the initial integral condition 2.1 (with $%
n=0)$ satisfied. Then there exists a number $\eta_{0}>0$ such that if 
\begin{equation}
\phi(t_{0})\equiv E_{tot}(t_{0})<\eta_{0}
\end{equation}
these Einstein equations have a solution on $M\times\lbrack t_{0},\infty),$
with initial values determined by $\sigma_{0},q_{0},u_{0},\overset{.}{u}%
_{0}. $ The parameter $t$ is $t=-\tau^{-1},$ with $\tau$ the mean extrinsic
curvature of $\Sigma\times\{t\}$ in the lorentzian metric $^{(3)}g$ on $%
\Sigma \times\lbrack t_{0},\infty).$

This solution is unique\footnote{{\footnotesize The global uniqueness
theorem of CB-Geroch says that it is geometrically unique in the class of
globally hyperbolic spacetimes.}} up to the choice of a section of
Tecichmuller space and a gauge choice for $A.$
\end{theorem}

\begin{proof}
We first prove that $E_{tot}(t)$ is uniformly bounded, and decays to zero
(without a priori hypothesis). We have obtained in the previous sections,
under the hypotheses H$_{E}^{\eta}$ and H$_{\sigma},$ the following result:
there are numbers $M_{i}$ depending only on $c_{E}$ and $c_{\sigma}$ such
that 
\begin{equation}
t^{k}E_{tot}(t)\leq M_{2}E_{tot}(t_{0})
\end{equation}
and 
\begin{equation}
|Q(t)-Q(t_{0})|\leq M_{3}\phi(t_{0})^{1/2}.
\end{equation}
Now consider the pair of $t$ dependent numbers 
\begin{equation*}
(\phi(t),\zeta(t)),\text{ \ \ }\zeta(t)\equiv|Q(t)-Q(t_{0})|
\end{equation*}
The inequalities 13.3, 13.4 show that the hypothesis (where $c_{E}$
satisfies H$_{E}^{\eta})$%
\begin{equation*}
\text{\ }\phi(t)\leq c_{E},\text{ }\zeta(t)\leq c_{\sigma}
\end{equation*}
imply that there exists $\eta_{0}>0$ such that $\phi(t_{0})\leq\eta_{0}$
implies that the pair belongs to the subset $U_{1}\subset R^{2}$ defined by
the inequalities: 
\begin{equation*}
U_{1}\equiv\{\text{\ }\phi(t)<c_{E},\text{ }\zeta(t)<c_{\sigma}\}.
\end{equation*}
Therefore for such an $\eta_{0}$ the pair belongs either to $U_{1}$ or to
the subset $U_{2}$ \ defined by 
\begin{equation*}
U_{2}\equiv\{\text{ }\phi(t)>c_{E}\text{ \ \ or \ }\zeta(t)>c_{\sigma}\}
\end{equation*}
These subsets are disjoint. We have supposed that for $\ t=t_{0}$ it holds
that $(\phi(t_{0}),\zeta(t_{0}))\in U_{1}$ hence, by continuity in $t$, $($\ 
$\phi(t),$ $\zeta(t))\in U_{1}$ for all $t$.

We have now proved that the total energy is uniformly bounded, and $\sigma
_{t}$ uniformly equivalent to $\sigma_{0}.$

To complete the proof of existence of the spacetime for $t\in\lbrack
t_{0},\infty)$ we need the following lemma.

\textbf{Lemma. }The $H_{2}$ norm of the pair of scalar functions ($%
\gamma,\omega)$ is uniformly bounded, as well of the $H_{1}$ norm of $%
(\partial_{t}\gamma,\partial_{t}\omega).$

\textbf{Proof of lemma. }We have already proven in section 9 the uniform
bound of \TEXTsymbol{\vert}\TEXTsymbol{\vert}$D\gamma||$ and $%
||D^{2}\gamma|| $ in terms of the total energy. On the other hand it holds
that 
\begin{equation}
\gamma-\gamma_{0}=\int_{t_{0}}^{t}\partial_{t}\gamma dt
\end{equation}
hence 
\begin{equation}
||\gamma-\gamma_{0}||\leq\int_{t_{0}}^{t}||\partial_{t}\gamma||dt
\end{equation}
Using previous estimates, the fall off of the energy and the property 
\begin{equation}
||\partial_{t}\gamma||\leq e^{-\lambda_{m}}||\partial_{t}\gamma||_{g}
\end{equation}
we find that there exists a number $M$ such that 
\begin{equation}
||\gamma-\gamma_{0}||\leq M\int_{t_{0}}^{t}t^{-(1+k)}dt,
\end{equation}
which completes the proof of the uniform bound of $||\gamma||_{H_{2}},$
hence also of $\gamma$ in $C^{0}.$

When $\gamma_{M}$ is uniformly bounded one can bound $||D\omega||\leq
e^{2\gamma_{M}}||e^{-2\gamma}D\omega||$ with the first energy and $%
||D^{2}\omega||$ with the total energy, in a manner analogous as the one
used for $||D^{2}\gamma||.$ We just recalled the estimate of \TEXTsymbol{%
\vert}\TEXTsymbol{\vert}$\partial_{t}\gamma||,$ the estimate of $%
||\partial_{t}\omega||$ is similar, when $\gamma$ has been bounded. It is
also easy to bound $||D\partial _{t}\omega||$ and $||D\partial_{t}\gamma||.$
\end{proof}

\begin{corollary}
1.\ This solution is globally hyperbolic, future timelike and null complete.

2. It is asymptotic to a flat solution: 
\begin{equation}
^{(4)}g=-4dt^{2}+2t^{2}\sigma_{\infty}+\theta_{\infty}^{2}
\end{equation}
with $\sigma_{\infty}$ a metric on $\Sigma$ independent of $t$ and of scalar
curvature $-1$, and $\theta_{\infty}$ a 1-form on $\Sigma\times S^{1}$ of
the type 
\begin{equation}
\theta_{\infty}=C(dx^{3}+H),
\end{equation}
where $C$ is a constant and $H$ is a harmonic 1-form on ($%
\Sigma,\sigma_{0}). $
\end{corollary}

\begin{proof}
1.\ The orthogonal trajectories to the space sections $M\times\{t\}$ have an
infinite proper length since the lapse $N$ is bounded below by a strictly
positive number. It can be checked that the conditions given in C.B and
Cotsakis for global hyperbolicity, and for future and null completeness are
satisfied by $(\Sigma\times R,^{(3)}g)$.

2. The theorem 5.7 and the decay of $\varepsilon+\varepsilon_{1}$ show that $%
\lambda$ tends to $2t^{2}$ in $C^{0}$ norm when $t$ tends to infinity.

The decay estimate of $\frac{dQ}{dt}$ show that $Q$ tends to a point $%
Q_{\infty}$ in $T_{eich}$ when $t$ tends to infinity, $\sigma$ tends to $%
\sigma(Q_{\infty}).$

The lapse and shift estimates 6.2 and 7.9, 7.11 show that $N$ tends to 2 and 
$\nu$ tends to zero in $C^{0}$ norm when $t$ tends to infinity.

The integral formula for $\gamma$ shows that $\gamma(t,.)-\gamma_{0}(.)$
tends to a function on $\Sigma,$ $\hat{\gamma}_{\infty}(.),$ in $L^{2}$ norm
when $t$ tends to infinity, hence $\gamma$ tends to $\gamma_{\infty}=\gamma
_{0}+\hat{\gamma}_{\infty}$ in this norm, therefore a fortiori $\gamma(t,.)$
tends to $\gamma_{\infty}(.)$ in the sense of distributions on $\Sigma$. We
know on the other hand that \TEXTsymbol{\vert}\TEXTsymbol{\vert}$D\gamma||$
tends to zero, hence $D\gamma$ tends to zero in the sense of distributions.
Since derivation in this sense is a continuous operator it holds that $%
D\gamma_{\infty}=0,$ therefore $\gamma_{\infty}$ is a constant.

An analogous reasonning holds for $\omega.$ The value of $\omega_{\infty}$
does not appear in the expression of $F.$

The estimates of section 2.2\ of $\hat{A}$ and $A_{t}$ (in Coulomb gauge)
show that they both tend to zero in $C^{0}$ norm on $\Sigma$. The
differential formula giving the $c_{i}(t)$ shows then that the 1 form $%
\tilde{A}$ tends in $C^{0}$ norm to the harmonic form on $\Sigma,$ $%
H_{\infty}=c_{i,\infty}H_{(i)}$The spacetime metric is asymptotic to the
metric 
\begin{equation}
^{(4)}g=e^{-2\gamma_{\infty}}(-4dt^{2}+2t^{2}\sigma_{\infty})+e^{2\gamma
_{\infty}}(dx^{3}+H_{\infty})^{2}
\end{equation}
which takes the indicated form by rescaling of $t.$
\end{proof}

\textbf{Aknowledgements. }We thank the University of the Aegean in Samos,
the Schrodinger Institute in Vienna and the Institut des Hautes Etudes
Scientifiques in Bures sur Yvette which made fruitful discussions with V.
Moncrief possible.

\textbf{References}

[A-M-T] L.\ Andersson, V.\ Moncrief and A. Tromba On the global evolution
problem in 2+1 gravity J.\ Geom. Phys. 23 1997 n$%
{{}^\circ}%
3-4$,1991-205

[CB1] Y.\ Choquet-Bruhat Global wave maps on curved spacetimes, in
''Mathematical and Quantum Aspects of Relativity and Cosmology'' Cotsakis
and Gibbons ed. LNP 535, Springer 1998, 1-30

[CB2] Y.\ Choquet-Bruhat Wave maps in General Relativity in ''On Einstein
path'' A. Harvey ed. Springer 1998, 161-185.

[CB-C] Y.\ Choquet-Bruhat and S. Cotsakis Global hyperbolicity and
completeness, J.\ Geom. Phys. 43 $n%
{{}^\circ}%
4$, 2002, 345-350.

[CB-DM] Y.\ Choquet-Bruhat and C DeWitt-Morette Analysis Manifolds and
Physics II enlarged edition (2000).

[CB-M 1]\textbf{\ }Y.\ Choquet-Bruhat and V.\ Moncrief, Future global in
time einsteinian spacetimes with U(1) isometry group Ann. Henri.
Poincar\'{e} 2 (2001), 1007-1064.

[CB-M 2]\textbf{\ }Y.\ Choquet-Bruhat and V.\ Moncrief, Non linear stability
of einsteinian spacetimes with U(1) isometry group, in Partial differential
equations and mathematical physics, in honor of J. Leray, Kajitani and
Vaillant ed. to appear 2003, Birkahauser.

[CB-M 3] Y.\ Choquet-Bruhat and V.\ Moncrief Existence theorem for solutions
of Einstein equations with 1 parameter spacelike isometry group, Proc.
Symposia in Pure Math, 59, 1996, H.\ Brezis and I.E.\ Segal ed. 67-80.

[CB-Y] Y.\ Choquet-Bruhat and J.\ W.\ York Geometrical well posed system for
the Einstein equations, C. R.\ Acad. Sciences Paris 321, 1995 1089-1095.

[F-T] A.\ Fisher and A. Tromba Teichmuller spaces, Math Ann. 267 1984,
311-345.

[M1] V.\ Moncrief Reduction of Einstein equations for vacuum spacetimes with
U(1) spacelike isometry group, Annals of Physics 167 (1986), 118-142

[M2] V. Moncrief Reduction of the Einstein - Maxwell and the Einstein -
Maxwell - Higgs equations for cosmological spacetimes with U(1) spacelike
isometry group. Class. Quantum Grav. 7 (1990) 329-352.

[M-S] S.\ Muller and M.\ Struwe Global existence of wave maps in 2+1
dimensions with finite energy data, Top. Met. in non lin. An. 7, n$%
{{}^\circ}%
2$ 1996, 245-261.

{\footnotesize YCB Universit\'{e} Paris 6, 4 Place Jussieu 75232 Paris
France Email YCB@CCR.jussieu.fr}

\end{document}